\documentclass{aa}
\usepackage{graphicx,times}

\begin{document}

\title{XMM-Newton confirmation of Soft X-ray excess emission 
in clusters of galaxies
 --
the discovery of O VII emission from an extended warm baryonic component}

\author{ J.S. Kaastra \inst{1}
	 \and
	 R. Lieu \inst{2}
         \and
         T. Tamura \inst{1}
         \and
         F.B.S. Paerels \inst{3}
         \and
         J.W. den Herder \inst{1}
         }
  
\offprints{J.S. Kaastra}
\mail{J.Kaastra@sron.nl}

\institute{ SRON National Institute for Space Research
              Sorbonnelaan 2, 3584 CA Utrecht, The Nether\-lands 
              \and
              Department of Physics, University of Alabama 
              in Huntsville, Huntsville, AL 35899
              \and
              Department of Astronomy, Columbia University, 
              550 West 120th Street, New York, NY 10027,
              }

\date{Received  / Accepted  }

\abstract{
We investigate a sample of 14 clusters of galaxies observed with XMM-Newton in a
search for soft X-ray excess emission.  In five of these clusters a significant
soft excess is evident.  This soft X-ray excess is compared with the thermal
emission from both the hot intracluster gas and any cooling (flow) gas that may
be present.  A warm (k$T=$0.2~keV), extended (several Mpc), plasma component is
particularly clear in the outer parts of the cluster, where the normal cluster
X-ray emission is weak.  This warm component causes both a thermal soft X-ray
excess at low energies (below 0.4--0.5~keV), as well as \ion{O}{vii} line
emission with a redshift consistent with a cluster origin, and not easily
interpreted as Galactic foreground emission.  The intensity of this component is
commensurate with what has been measured before with the ROSAT PSPC in the
1/4~keV band.  We attribute this component to emission from intercluster
filaments of the Warm-Hot Intergalactic Medium in the vicinity of these
clusters.  For the central regions of clusters the detection of lines in the
soft X-ray spectrum is more difficult, due to the predominance of the X-ray
emitting hot plasma there, hence we cannot discriminate between the thermal and
nonthermal origin of the soft excess, leaving several options open.  These
include thermal emission from warm filaments seen in projection in front of or
behind the cluster center, thermal or nonthermal emission in the cluster core
itself related to magnetic reconnection, or Inverse Compton emission from the
cosmic microwave background on relativistic electrons.
\keywords{Galaxies: clusters: general --
-- Galaxies: cooling flows --
X-rays: Galaxies: clusters }}

\titlerunning{XMM-Newton   Soft X-ray excess  
in clusters of galaxies}

\maketitle

\section{Introduction}

The phenomenon of {\it soft excess} in clusters of galaxies, viz.  diffuse soft
X-ray and EUV radiation at a level higher than that expected from the hot
intracluster medium, has been a subject of controversy ever since the time of
the first reports (Lieu et al.  \cite{lieu1996a}, \cite{lieucoma}, Mittaz et al.
\cite{mittaz98}).  As a result, a large number of papers on both theoretical and
observational aspects appeared.  On the former, in spite of the many ideas
proposed, there are only two prevailing interpretations of the phenomenon:  (a)
the original thermal warm plasma emission model from the three papers cited
above (see also Fabian \cite{fabian1997}, and Cen \& Ostriker \cite{cen1999});
(b) inverse-Compton scattering between the cosmic microwave background (CMB) and
intracluster relativistic electrons (Hwang \cite{hwang1997}, Ensslin \& Biermann
\cite{ensslin1998}, Sarazin \& Lieu \cite{sarazin1998}).  On the latter, an
important feature is the soft excess surface brightness relative to that of the
hot virialized thermal component - this in general was seen to rise with cluster
radius.  The behavior was first realized in A~1795 (Mittaz et al.
\cite{mittaz98}) and subsequently found to be conspicuously present also in
A~2199 (Lieu et al.  \cite{lieu1999a}), where on both occasions it was taken as
evidence for pre-virialized matter falling into the cluster while getting
heated.  Later came results for the Shapley supercluster region where both the
ROSAT PSPC and BeppoSAX LECS spectra exhibit the same effect (Bonamente et al.
\cite{bonamente2001a}).  Most recently, a detailed analysis of the EUVE data of
five clusters using wavelet reconstruction technique (Durret et al.
\cite{durret}), and a synoptic study of 38 X-ray bright clusters in directions
of low Galactic absorption (Bonamente et al.  \cite{bonamente2002}), revealed
the rising trend of the soft excess as commonly occurring, and points to the
existence of a soft radiation component from the outer region of clusters where
there is insufficient virialized plasma to confine any relativistic particles.
The radiation from such regions is likely to be signature of warm filaments
where a significant number of baryons is expected to reside (Cen \& Ostriker
\cite{cen1999}, Bonamente et al.  \cite{bonamente2002}).

Owing partly to the challenges presented by the soft excess phenomenon to our
understanding of clusters and large scale structures, papers also appeared which
questioned the reality of the phenomenon in many of the cases described above.
Key examples include the work of Bowyer et al.  (\cite{bowyer1999}) which
attributed the EUV excess at large cluster radii of A~1795 and A~2199 to an
artifact of erroneous background subtraction, and Arabadjis \& Bregman
(\cite{arabadjis1999}) who reported the absence of a soft X-ray excess in many
clusters observed by the ROSAT PSPC when these data were modeled with
interstellar absorption correctly taken into account.  The former was addressed
by a dedicated re-observation of A~1795, A~2199, and Virgo using for the purpose
of subtraction an {\it in situ} background measurement obtained by pointing at
small offset from the cluster (Lieu et al.  \cite{lieu1999a}, Bonamente et al.
\cite{bonamente2001b}); in each case the EUV excess was found to remain present
at statistically significant levels for the regions in question.  The latter may
be rebutted by noting that (a) when the ROSAT PSPC data of many clusters were
analyzed or re-analyzed using the correct absorption code by the authors
mentioned in the previous paragraph, a soft X-ray excess was discovered; (b)
also pointed out earlier is the radially rising importance of the soft component
which in the context of Arabadjis \& Bregman (\cite{arabadjis1999}) would imply
a cosmic conspiracy - somehow our interstellar gas distribution is correlated
with the central position of many clusters.

Here we investigate the soft excess phenomenon using data obtained by the EPIC
cameras (Str\"uder et al.  \cite{strueder}; Turner et al.  \cite{turner})
onboard the XMM-Newton satellite.  Compared with the ROSAT PSPC or BeppoSAX LECS
instruments, the EPIC cameras have a higher sensitivity, better spatial and
spectral resolution and cover a broader energy range.  Reliable spectra can be
obtained for energies larger than $\sim$0.2~keV.  Only EUVE is sensitive to even
softer photons, but this instrument lacks spectral resolving power.

\section{Data selection}

\subsection{Cluster sample}

In this paper we examine the same sample of 14 clusters described in more detail
by Kaastra et al.  (\cite{kaastra2002}).  Briefly, the sample consists of 13
clusters with strong cooling flows, taken from the Guaranteed Time program of
the RGS consortium and other available data sets.  These cooling flow clusters
were selected mainly based upon their suitability for spectroscopic studies with
the Reflection Grating Spectrometers (RGS) of XMM-Newton.  This selection
criterion favours relatively medium distant, compact and cool clusters.  Nearby
clusters in general have a large angular size, resulting in a degradation of the
spectral resolving power of the RGS.  In very hot clusters the line emission is
weak with respect to the Bremsstrahlung continuum.  The Coma cluster was added
as a control case of a cluster without cooling flow.  The cluster sample is
listed in Table~\ref{tab:sample}.

\begin{table}[!h]
\caption{Cluster sample, sorted in order of increasing Galactic
column density (Table~\ref{tab:nh3}). The net exposure times are the
weighted average of the MOS1, MOS2 and pn camera's, with weights
0.25, 0.25 and 0.5, respectively.}
\label{tab:sample}
\centerline{
\begin{tabular}{|lrrr|}
\hline
Cluster & redshift & Orbit & exposure \\
        &          & nr.   & (ks)     \\
\hline
\object{Coma}              & 0.0240 &  86 & 14 \\
\object{A 1795}            & 0.0631 & 100 & 26 \\
\object{A 4059}            & 0.0475 & 176 & 20 \\
\object{S\'ersic~159-03}   & 0.0580 &  77 & 30 \\
\object{Virgo}             & 0.0043 &  97 & 29 \\
\object{A 1835}            & 0.2532 & 101 & 25 \\
\object{MKW 3s}            & 0.0450 & 129 & 35 \\
\object{A 2052}            & 0.0350 & 128 & 27 \\
\object{NGC 533}           & 0.0185 & 195 & 35 \\
\object{A 496}             & 0.0324 & 211 &  7 \\
\object{A 1837}            & 0.0698 & 200 & 44 \\
\object{Hydra A}           & 0.0539 & 183 & 10 \\
\object{A 262}             & 0.0163 & 203 & 20 \\
\object{2A 0335+096}       & 0.0349 & 215 & 14 \\
\hline\noalign{\smallskip}
\end{tabular}
}
\end{table}

\subsection{Event selection and background subtraction}

The event selection and further analysis is described in more detail in another
paper (Kaastra et al.  \cite{kaastra2002}).  Here we give a short summary.  For
most of the clusters, the data were obtained in the full frame mode, with the
thin filter.  Exceptions are A~496, A~1837 and A~2052 (extended full frame for
pn), and Coma (medium filter).  For A~1835 we discarded the MOS1 data (taken in
a large window mode).  The data were extracted using the standard SAS software,
equivalent to version 5.3.  For MOS, we included events with $0\le$pattern$\le
12$, for pn we took only pattern$=0$.  Not all hot pixels get removed by the
standard SAS procedures.  In particular at low energies this happens sometimes.
We have verified, however, that hot pixels do not contribute significantly to
the soft events.

The background spectrum of the XMM-Newton EPIC camera's is described in detail
by Lumb et al.  (\cite{lumb}).  The main components are the cosmic X-ray
background, dominating the background at low energies, and a time-variable
particle component, dominating the high energy background.

\subsubsection{Time-variable particle background}

The time-variable particle component is caused by clouds of soft protons.  Apart
from long quiescent intervals, there are sporadic episodes where this background
component is enhanced by orders of magnitude, as well as time intervals with
only a low level of flaring.

The contribution of X-ray photons to the full field 10--12~keV band count rate
is almost negligible as compared with the background.  This energy band is
dominated by low energy protons.  Therefore it can be used as a monitor for the
time-variable background component.  Thus we divided the observation into time
intervals of 260~s, and intervals with high particle background (as measured in
the 10--12~keV band) were excluded.  The threshold level for rejection was set
to 35 counts / 260~s for MOS and 50 counts / 260~s for pn.  This threshold is
about 2.5 and 3.5$\sigma$ above the average quiescent level of 23 and 31 counts
/ 260~s for MOS and pn, respectively.  Only in a few cases with enhanced
background did we adopt a slightly higher threshold than the above.  This was
the case in A~4059, A~496, Hydra~A and 2A~0335+096, where the threshold level
was set to 50 for MOS and 56 for pn.

The above procedure is sufficient to remove the large flares, simply by
discarding the time intervals with high 10--12~keV count rates.  However after
removing these intense large solar proton events the average background level
still varies slightly.  The r.m.s.  variation from cluster to cluster is
$\sim$10--20~\%.  This is due to the fact that for some observations the count
rate is slightly enhanced but smaller than the cutoff threshold.  This holds in
particular for the clusters with a higher threshold but also to some extent for
the others.  These remaining weak flares cannot be removed by time selection.
For them, however, the increase in background flux in the 0.2--10~keV raw
spectrum is to first order proportional to the increase in 10--12~keV count
rate.  We utilize this behaviour to make a first order correction to the
estimated background spectrum.  The usual procedure would be to subtract the
background spectrum using a blank field observation screened in the same way as
the cluster data.  Here in order to correct for the weak variability we divided
the cluster and blank field observation into subsets with the same 10--12~keV
count rate.  Using these subsets we normalized the blank field background to
have a flux distribution in the 10--12~keV band similar to the cluster
observation.  This ensures correct particle background subtraction, under the
assumption that for low proton count rates the shape of the proton spectrum does
not vary when its flux increases by a small amount.  However in order to account
for any possible remaining particle background subtraction problems, we have
included a systematic uncertainty of at least 10~\% of the total background in
all our fits.  Because the background at low energies is dominated by diffuse
cosmic X-rays, and in most cases the low energy flux of the clusters we study is
sufficiently well above the total (particle plus X-ray) background, our results
are not very sensitive to these details in the background.

The deep field used in our background subtraction procedure consisted of the
background event file provided by the XMM-Newton Science Operations Center,
containing a total of 320--410~ks exposure time on 8 deep fields (the exposure
time differs slightly for the different instruments).  This background event
file was filtered the same way as the source file.

As stated above, the contribution of X-ray photons to the 10--12~keV band count
rate is almost negligible.  Only for Coma, the hottest and brightest cluster in
our sample, the full-field 10--12~keV cluster emission reaches a level of 10~\%
of the quiescent background count rate in the pn camera only.  Owing to the
lower sensitivity of the MOS cameras at high energies, the relative cluster
contamination in the 10--12~keV band is at most 5~\% there.  Thus, we expect
that our method over-estimates the time-variable particle contribution in Coma
by 5--10~\%.  This does not affect our science results for Coma, however, since
(i) the difference is within the systematic background uncertainty used by us;
(ii) at low energies the background is dominated by the constant cosmic X-ray
background, and not by the soft proton component; (iii) except for the outermost
annulus, the X-ray flux of Coma is a factor of 5--100 times brighter than the
subtracted background for all energies below $\sim$6~keV; (iv) the spectral fit
is dominated by the higher signal-to-noise part of the spectrum below
$\sim$6~keV.

\subsubsection{Cosmic X-ray background}

While at high energies the particle contribution dominates the background
spectrum, at low energies the cosmic X-ray background yields the largest
contribution to the measured background.  The cosmic X-ray background varies
from position to position on the sky.  This is in particular important at low
energies.  To get a typical estimate for the sky variation, we took the PSPC
count rates for 378 regions at high Galactic latitude as studied by Snowden et
al.  (\cite{snowden2000}) and determined the population variation for this
component in the R1 (low energy) band; the r.m.s.  variation is typically
35$\pm$3~\%.  In principle, the background estimate for any location on the sky
can be improved by taking the measured PSPC count rates into account.  However,
this requires detailed spectral modeling of the soft X-ray background for each
position on the sky that is being studied, which can be rather uncertain due to
the poor spectral resolution of the PSPC.  Here we have taken a conservative
approach by using the average background as contained in the standard EPIC
background files, but we included a systematic background error of 35~\% of the
total background below 0.5~keV, 25~\% between 0.5--0.7~keV, 15~\% between
0.7--2~keV, and 10~\% above 10~keV.  This last 10~\% also includes the
systematic uncertainty in the subtracted time-variable particle background
component.

Note that the above 35~\% systematic background uncertainty at low energies is a
r.m.s.  estimate.  In individual cases there may be a larger deviation from the
average high latitude X-ray background.

The differences between the spectral response for pn data taken with the full
frame mode (as in our deep fields) and the extended full frame mode (as in some
of our clusters) are negligible compared to the systematic uncertainties on the
background and effective area that we use in this paper.  Of more importance is
the difference between data taken with the medium filter (Coma) and the thin
filter (all other clusters and the deep fields).  For a power law photon
spectrum with photon index 2, the medium filter produces about 10~\% less counts
at 0.2~keV, 6~\% less at 0.3~keV, and less than 2~\% above 0.4~keV.  These are
all well within our adopted systematic uncertainties.

Finally, in all fields the brightest X-ray point sources were removed.

\subsection{Assessment of the calibration accuracy\label{sec:cal}}

Since soft excess emission occurs near the lower boundary of the useful energy
range of the EPIC camera's, assessing the current accuracy of the calibration is
an important issue.  Currently the quoted accuracies for the EPIC calibrations
of the effective area are in the order of 5~\% between 0.4 and 10 keV for pn
(Briel et al.  \cite{briel}) and also a 5~\% between 0.2 and 10 keV for MOS
(Ferrando et al.  \cite{ferrando}).  A larger difference exists between the pn
and MOS cameras at higher energies ($>$5 keV).

Hence we have investigated the quality of the calibration specifically for the
low energy part of the spectrum.  We have studied MOS2 spectra, taken with the
thin filter, of the BL Lac object PKS~2155-304.  The thin filter is used for all
our clusters except for Coma.  The spectrum of PKS~2155-304 is known to be a
power law over a broad energy range, which is confirmed by high resolution
grating spectra (XMM-Newton RGS, Chandra LETGS).  An analysis by our group of
the LETGS data of PKS~2155-304 leads to a best estimate of the Galactic column
density of $(1.27\pm 0.02)\times 10^{20}$~cm$^{-2}$, within the error bars of
the independently measured 21~cm column density ($(1.36\pm 0.10)\times
10^{20}$~cm$^{-2}$, Lockman \& Savage \cite{lockman}).

Fitting this spectrum over the full 0.2--10~keV range (as we will do for our
clusters) with a power law and keeping the absorption column fixed to the above
value, leads to an acceptable fit.  In the low energy range below 1~keV the fit
residuals are in general less than 5~\%, while the fit shows an apparent soft
excess in the 0.2--0.3~keV band of 5~\%.  This excess should be attributed to
the remaining calibration uncertainties.  If instead we leave the column density
as a free parameter, the best fit value is $0.87\times 10^{20}$~cm$^{-2}$, i.e.
$0.40\times 10^{20}$~cm$^{-2}$ smaller than the Galactic value.  The fit
residuals over most of the energy range have only slightly changed, and are
still less than 5~\% below 1~keV.  The "soft excess" in the 0.2--0.3~keV band
has diminished now from 5~\% to a negligible 2~\%.

We conclude from the above that in spectral fits over the full 0.2--10~keV
range, a 5~\% excess in the 0.2--0.3~keV band (or any other higher energy band)
falls just within the calibration uncertainty, and an apparent column density
deficit of $0.40\times 10^{20}$~cm$^{-2}$ is also within this uncertainty range.

We have repeated the same analysis using pn data of PKS~2155-304.  Typically the
fit residuals are a factor of 1.4 larger as compared with MOS2.

Hence, in the following we will only consider detections of a soft excess secure
for those spectra where the measured excess corresponds to 10~\% or more in the
0.2--0.3~keV band, or apparent column density deficit larger than $0.80\times
10^{20}$~cm$^{-2}$).

Combining MOS and pn data, we have accommodated for these uncertainties by using
systematic errors to the spectra of 10~\% of the source flux below 0.3~keV and
above 2~keV, and 5~\% in the 0.3--2~keV range.  This is in addition to the
systematic uncertainty in the background.

\subsection{Spectral extraction and fitting}

Spectra were extracted in 9 annuli with outer radii of 0.5, 1, 2, 3, 4, 6, 9, 12
and 16\arcmin.  The effective area for each annulus was determined as follows.
First, a template image was created (for which we took a vignetting-corrected
ROSAT PSPC image, obtained from NASA's skyview utility).  This image was then
corrected for energy-dependent vignetting (both due to the telescope and in case
of the MOS cameras also due to the Reflection Grating Array).  This vignetted
image was convolved with the energy and position dependent telescope point
spread function .  After this the exposure correction for bad pixels/columns,
CCD gaps and excluded point sources was taken into account.  This procedure was
followed for a grid of energies.  The ratio of the number of counts in these
simulated images to the number of counts in the original image was then used as
the (energy-dependent) correction factor to the full on-axis effective area.  No
deprojection corrections were applied.  The spectra for each annulus were then
fitted individually.  Again, more details and justification for this procedure
are given by Kaastra et al.  (\cite{kaastra2002}).

The template PSPC images represent the low-energy emssion ($E<2.5$~keV).
However, this does not strongly affect our effective area correction for high
energies, because the image mainly serves to estimate what fraction of the
photons from a given annulus are detected in the neighbouring annuli due to the
instrumental point spread function of XMM-Newton.  This fraction depends only a
little upon the shape of the image within the annulus, not upon its
normalisation.  Our method does not produce bias as long as the spectral shape
does not vary on a smaller scale compared with the width of the annulus.  This
condition is met in all of our clusters.

The spectra for each annulus were accumulated in 15~eV bins.  These bins were
then binned further onto a grid with spacing of about 1/3~FWHM for all three
detectors.  We used the same energy grid for all three detectors.  Further
binning with a factor between 2--12, intended to enhance the S/N ratio for the
weak spectra of the outer parts, was performed for all spectra, in particular
for the parts of the spectrum where no spectral lines are present.  Spectral
fitting was restricted to the 0.2--10~keV range.  The energy range below 0.2~keV
is currently too poorly calibrated to be useful for spectral analysis.

For the spectral analysis, we used the SPEX package (Kaastra et al.
\cite{kaastra2002spex}).  For the interstellar absorption, we used the Morrison
\& McCammon (\cite{morrison}), and for thermal plasma emission we used the
collisional ionization equilibrium (CIE) model as available in SPEX.  Systematic
errors, both as a fraction of the source spectrum and the background spectrum,
were added according to the prescriptions of the previous sections.

We fitted all three EPIC spectra together.  The MOS and pn fit results are
consistent with each other, as follows from plots of the fit residuals that we
made for each spectrum.  There were no significant differences larger than the
systematic uncertainties assigned to each energy, and for our best-fit models
the $\chi^2$ values showed full consistency with the model for most spectra.

\section{Spectral analysis}

\subsection{Basic spectral model\label{sec:basicmodel}}

As a basic spectral model, we use at each radius a two temperature model with
Galactic absorption.  The abundances are left as free parameters.  In the center
of most of our clusters (except for Coma) there is a combination of hot and
cooler gas.  As RGS spectra of all our clusters have shown (Peterson et al.
\cite{peterson2001}; Kaastra et al.  \cite{kaastra2001}; Tamura et al.
\cite{tamura1795}, \cite{tamura496}, Peterson et al.  \cite{peterson2002}), the
amount of cooling gas in the cluster centers is much smaller than predicted by
the standard isobaric cooling flow model, and in most clusters there is not much
gas found below a temperature of about a third of the hot gas temperature.
Therefore we decided to constrain the model to a two temperature model, with the
temperature of the coolest gas fixed to half the temperature of the hot gas.
This recipe keeps the fitting procedure stable.  In fact, a factor of about two
difference in temperature between both components is just the maximum
temperature resolution that can be reached for any spectrum.  It corresponds to
the typical temperatures of which most ions have a significantly different
concentration.  With CCD spectral resolution, as we have here, the temperature
structure is mostly constrained by the different line centroids of the L-shell
spectra of iron ions, each ion being sensitive to a different temperature range.
In the fit we left the abundances of Fe, Si and O free, and since the strongest
Ne and Mg lines occur in the Fe-L shell energy band, which is the strongest
temperature indicator, we also left those abundances free.  In order not to
increase the number of free parameters unnecessarily, we coupled the abundances
of S, Ar, Ca and Ni to the Si abundance, and the abundances of C and N to the
abundance of O, using solar ratios (Anders \& Grevesse \cite{anders}).  However
our results are not too strongly dependent upon the precise abundances.  In
order to reduce the number of free parameters further, we fixed the abundances
of the cool component to those of the hot gas.

While a two temperature structure is the natural choice for the innermost
cooling part of the cluster, usually the gas in the outer parts of the cluster
is assumed to be isothermal, with a single temperature at each radius.  In some
clusters substructure in the outer parts may cause a broadened temperature
structure, and thus in those cases our two temperature approach will lead to a
better description.  Also in those cases where there are large scale variations
in temperature from one side of the cluster to the other, the annular spectra
may contain gas at different temperatures.  In those cases where the gas is
truly isothermal, we find that the emission measure of the coolest gas drops to
zero, and in several of our clusters this is indeed the case.

For each cluster we then determined the best fit two temperature model with
Galactic absorption as the starting point for further analysis.

In most of the clusters, the measured spectrum in the outermost, 9th annulus
(12--16\arcmin) is too noisy to yield useful constraints, mainly due to the
relatively large background in the EPIC data.  Therefore we restrict the
analysis to the innermost 8 annuli (radius less than 12\arcmin) for all our
clusters except for Coma and Virgo.  In addition, A~1835 has a relatively small
angular size (mainly due to its large redshift), and for that cluster we also
discard annulus 8 (9--12\arcmin).  The Coma annuli are centered on one of the
two dominant elliptical galaxies, NGC~4874, which has a distinct spectrum from
the rest of the cluster (Arnaud et al.  \cite{arnaud2001}); the same holds for
Virgo with M~87 (B\"ohringer et al.  \cite{boehringer2001}).  In both cases we
therefore discarded the innermost annulus ($<$0.5\arcmin) containing these
elliptical galaxies.  Finally, in A~4059 the standard two temperature model did
not provide a good fit.  Chandra observations of the central part of the cluster
(Heinz et al.  \cite{heinz}) show strong interaction between the central AGN and
the cluster medium with cavities etc.  Therefore we also discarded the innermost
annulus of A~4059 in our present analysis.

\subsection{Absorption deficit?}
\begin{figure}
\resizebox{\hsize}{!}{\includegraphics[angle=-90]{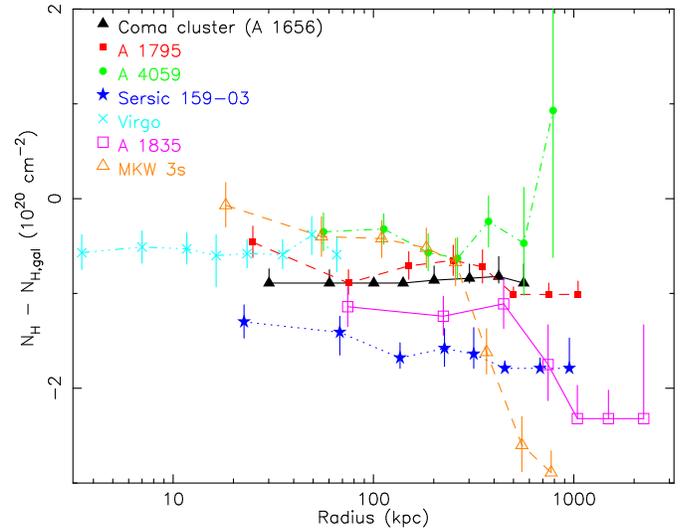}}
\caption{Difference of measured minus Galactic column density for 7
clusters of galaxies. Data points with no lower limit have a value of zero
for the measured Galactic column density.
}
\label{fig:nh2a}
\end{figure}
\begin{figure}
\resizebox{\hsize}{!}{\includegraphics[angle=-90]{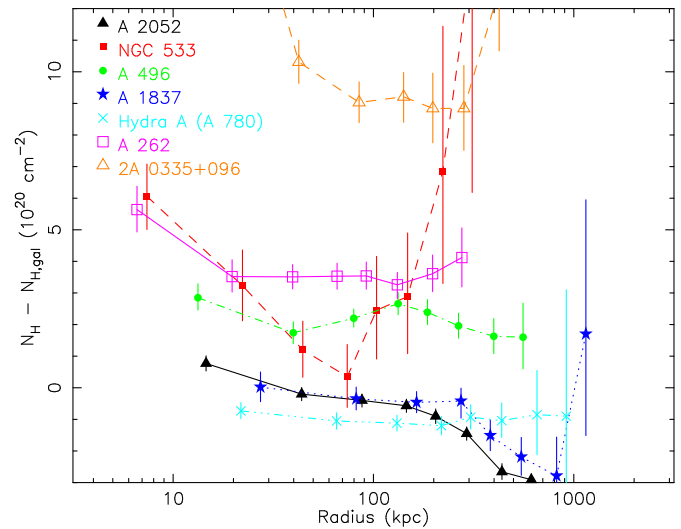}}
\caption{As Fig.~\ref{fig:nh2a}, but for 7 other clusters.
}
\label{fig:nh2b}
\end{figure}
As a first model to test the presence of soft excess emission in our clusters,
we have modified our standard two temperature model with fixed Galactic column
density.  Instead we allow the absorbing column density to be free.  We have
determined the best fit column densities and plot the difference with the
adopted Galactic column density in Figs.~\ref{fig:nh2a} and \ref{fig:nh2b}.

In most of our clusters the fitted Galactic column density appears to be almost
constant with radius.  In a few cases there appears to be a small relative
excess absorption in the innermost bin(s).  This may be real but could also be
due to inappropriate modeling of the complex cool gas distribution in the
innermost core, or modeling inaccuracies due to the strong spectral gradients in
the central bin, which has a size not much larger than the width of the
telescope point spread function.  Excess absorption in the cores of clusters of
galaxies will be studied in detail in another paper (Kaastra et al.  2002, in
preparation).

In a few clusters (MKW~3s, A~2052, A 1837) we observe a strong apparent drop of
the fitted column density beyond $200-300$~kpc radius.  In all these three
clusters this drop occurs beyond the radius where also the hot intracluster
medium starts to have a significantly lower temperature.

\begin{figure}
\resizebox{\hsize}{!}{\includegraphics[angle=-90]{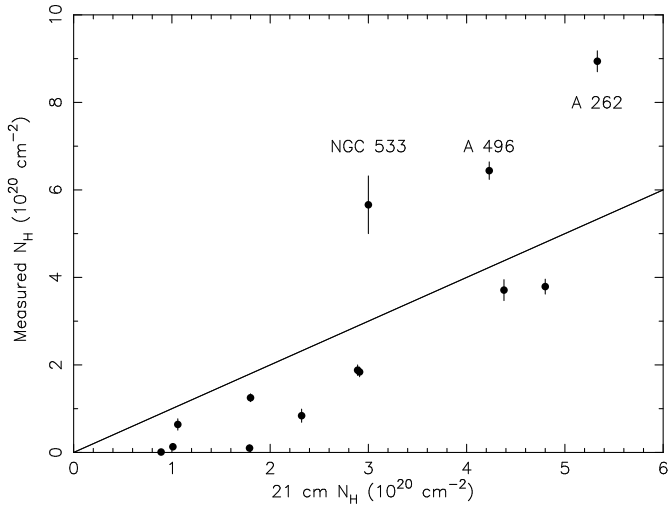}}
\caption{Measured versus Galactic column density. Cluster 2A~0335+096
with its high Galactic column density is not shown on the plot. The
other three clusters with excess absorption are indicated in the plot.}
\label{fig:nh3}
\end{figure}

We have also determined the weighted average of the column density for all our
clusters.  The results are summarized in Table~\ref{tab:nh3} and
Fig.~\ref{fig:nh3}.
\begin{table}[!h]
\caption{Column densities of the clusters, in units of $10^{20}$~cm$^{-2}$.}
\label{tab:nh3}
\centerline{
\begin{tabular}{|lrrr|}
\hline
Cluster & $N_{\mathrm{H,gal}}$ & $N_{\mathrm{H,obs}}$ 
 & $N_{\mathrm{H,obs}}-N_{\mathrm{H,gal}}$\\
\hline
{Coma}              &0.89 &0.01$\pm$ 0.05 &$-$0.88\\
{A 1795}            &1.01 &0.13$\pm$ 0.06 &$-$0.88\\
{A 4059}            &1.06 &0.64$\pm$ 0.13 &$-$0.42\\
{S\'ersic~159-03}   &1.79 &0.10$\pm$ 0.07 &$-$1.69\\
{Virgo}             &1.80 &1.25$\pm$ 0.09 &$-$0.55\\
{A 1835}            &2.32 &0.84$\pm$ 0.15 &$-$1.48\\
{MKW 3s}            &2.89 &1.88$\pm$ 0.12 &$-$1.01\\
{A 2052}            &2.91 &1.84$\pm$ 0.11 &$-$1.07\\
{NGC 533}           &3.00 &5.66$\pm$ 0.66 &$+$2.66\\
{A 496}             &4.23 &6.44$\pm$ 0.20 &$+$2.21\\
{A 1837}            &4.38 &3.71$\pm$ 0.24 &$-$0.67\\
{Hydra A}           &4.80 &3.79$\pm$ 0.17 &$-$1.01\\
{A 262}             &5.33 &8.94$\pm$ 0.24 &$+$3.61\\
{2A 0335+096}       &18.64&28.71$\pm$ 0.50&$+$10.07\\
\hline\noalign{\smallskip}
\end{tabular}
}
\end{table}

In 10 out of 14 clusters we obtain a lower column density than the Galactic
column density.  Only in four clusters we find excess absorption:  NGC~533,
A~496, A~262 and 2A~0335+096.  All these four clusters with apparent excess
absorption have rather poorly determined 21~cm column densities.  A higher total
column density of $(23\pm 4)\times 10^{20}$~cm$^{-2}$, consistent with what we
find here (($28.7\pm 0.5)\times 10^{20}$~cm$^{-2}$) was found before in PSPC
spectra of 2A~0335+096 (Irwin \& Sarazin \cite{irwin}).  Our column density in
A~262 is also consistent with an earlier PSPC analysis (David et al
\cite{david96}).  The higher column density in A~496 has also been reported
before by Tamura et al.  (\cite{tamura496}).

It is unclear why some of these column densities are so large.  For example, the
maps of Dickey and Lockman (\cite{dickey}) show a range of column densities
between $5.2-5.8\times 10^{20}$~cm$^{-2}$ for a cone of 2\degr\ around A~262,
clearly excluding a high value of $8.9\times 10^{20}$ as we measure here.
However, IRAS 100~$\mu$m maps (obtained with NASA's skyview facility) show
enhanced emission near the cluster and to the east of it, suggesting that
perhaps some additional absorption by Galactic dust and gas could explain the
excess absorption.  This was also suggested earlier by David et al.
(\cite{david96}).

A~496 is located at the NW boundary of a region of several degrees diameter with
enhanced infrared emission, so also here enhanced absorption is a legitimate
interpretation.

2A~0335+096 is located in a region of the sky which has already a large column
density, and the infrared maps show strong and patchy emission in this region.
So again excess absorption is not unexpected.

Only NGC~533 appears to be in a region with a rather smooth column density
distribution, with not much indication of enhanced infrared emission.
Inspection of Fig.~\ref{fig:nh2b} shows that the excess absorption is mainly due
to the innermost two annuli; we note that this cluster is very compact, in fact
its angular core radius is almost similar to that of A~1835, and its surface
brightness as observed by EPIC drops by an order of magnitude within the central
annulus.  Thus, similar to that cluster modeling problems related to the
instrumental psf in combination with the strong spectral gradients in the
central part may explain this deviation.

Therefore, a likely explanation for the apparent (NGC~533) or true (A~496, A~262
and 2A~0335+096) excess absorption in these four clusters can be given.  In the
remainder of this paper we therefore focus upon the 10 clusters with an
absorption deficit.  For consistency, we checked the IRAS 100~$\mu$m maps for
all these 10 clusters, and found only significant cirrus-like structure for
A~1837.  All others appeared to have rather smooth infrared maps.

For the 10 clusters with an apparent lower column density, the derived deficit
ranges between $0.4 - 1.7\times 10^{20}$~cm$^{-2}$.  The significance of this
deficit should be assesed against three sources of uncertainty:  the measured
error in the 21~cm column density, the statistical error in our X-ray
determination of the column density, and the systematic calibration uncertainty
of our instruments.

In most of these 10 clusters with an apparently lower column density, this
deficit is much larger than the typical uncertainty of about $0.1\times
10^{20}$~cm$^{-2}$ in the measured 21~cm column density.  For example, for Coma
the inferred column density is almost zero.  A total column density that is
significantly smaller than the Galactic column can thus be excluded.

The statistical errors on the X-ray column densities (Table~\ref{tab:nh3}, third
column), though already overestimated by the inclusion of systematic
uncertainties in our spectral modeling, are always smaller than $0.24\times
10^{20}$~cm$^{-2}$ for the 10 clusters with an absorption deficit.

The deficit in many of these clusters is also larger than the typical systematic
calibration uncertainty of $0.4\times 10^{20}$~cm$^{-2}$ (see
Sect.~\ref{sec:cal}).  The apparent lower column density is therefore most
likely due to excess emission instead of lack of absorption.  We conclude that
in several of our clusters there is a real soft excess.  In the next section we
argue that in five of the ten clusters with apparent absorption deficit the
phenomenon is due to a soft X-ray excess in the spectrum.  In the remaining
cases either the statistics are insufficient to establish a soft excess, or
there are other complicating factors.

\subsection{Significance of the soft excess}

In order to characterize the apparent soft excess and to assess its
significance, we have fitted the spectra using our previous model for a two
temperature plasma with Galactic absorption, plus in addition a third soft
emission component for which we consider three different cases:  a power law
spectrum, a thermal plasma with cluster abundances (coupled to the abundances of
the hot gas), and a thermal plasma with zero metallicity (pure H/He mixture).
\begin{figure}
\resizebox{\hsize}{!}{\includegraphics[angle=-90]{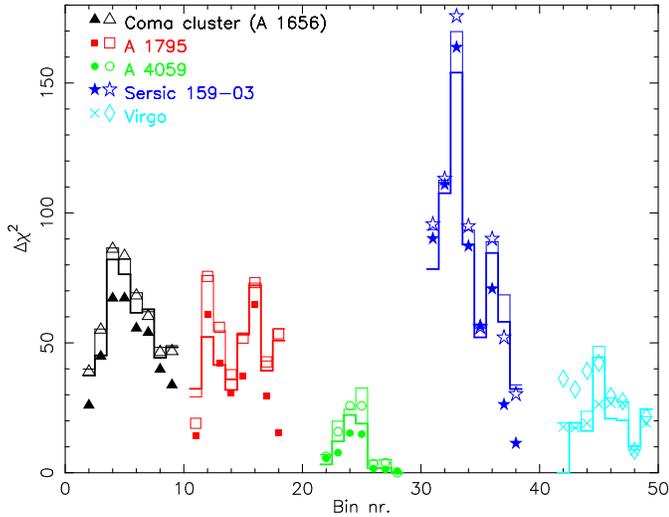}}
\caption{Improvement of $\chi^2$ for different models for the soft excess,
as compared with a fit with two temperatures and Galactic absorption.
Filled symbols: a lower than Galactic absorption column density;
Open symbols:
power law model; thick solid line: thermal plasma with cluster abundances;
thin solid line: thermal plasma with zero metallicity. For the
$k$th cluster the $\chi^2$ values for annulus $j$ are plotted at bin
number $10(k-1)+j$.}
\label{fig:chi1}
\end{figure}
\begin{figure}
\resizebox{\hsize}{!}{\includegraphics[angle=-90]{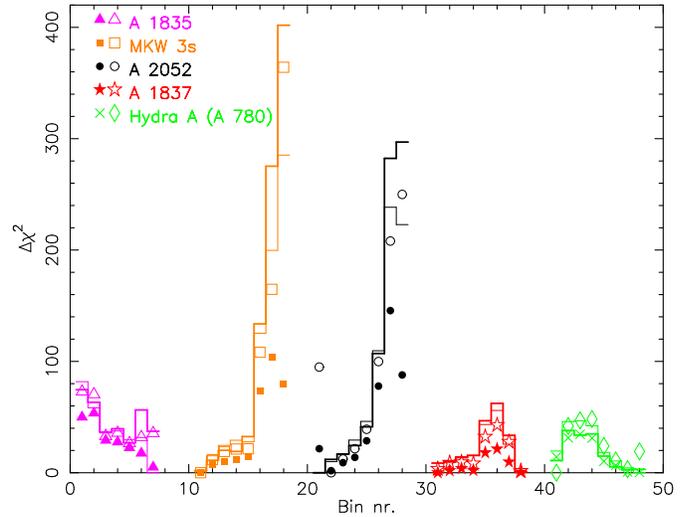}}
\caption{As Fig.~\ref{fig:chi1}, but for five other clusters.}
\label{fig:chi2}
\end{figure}

By introducing these additional soft emission components, the fits improve
dramatically.  This is shown in Figs.~\ref{fig:chi1} and \ref{fig:chi2}, where
the improvement in $\chi^2$ is shown as compared with the two temperature model
with Galactic column density.  All three models (as well as the model of a free
column density discussed in the previous section) give very similar improvements
in the fits.  How significant are the improvements in the fits?  We have
performed formal F-tests for all three models on all the cluster spectra.  A
significant improvement at the 95~\% confidence level or much higher is achieved
in at least two annuli for the following clusters:  Coma, A~1795,
S\'ersic~159--03, A~1835, MKW~3s and A~2052.  In these last two cases the excess
is in particular strong in the outermost annuli, while for S\'ersic~159--03 the
excess is strong for all radii.

In A~4059, Virgo, A~1837 and Hydra~A, the fit improvement is in general not very
significant.  In the first three cases also the apparent column density deficit
(Table~\ref{tab:nh3}) is within the suspicious range for systematic calibration
uncertainties (see section~\ref{sec:cal}).  Hydra~A has a rather large Galactic
column density, making the low energy flux small, and thereby the detection of
any soft component more difficult.

A~1835 is by far the most compact cluster in our sample, due to its large
redshift (0.25).  It only shows a significant excess in the innermost two
annuli, within a radius of 1\arcmin.  Possible small scale spectral gradients in
this innermost cooling core in combination with the instrumental point spread
function forces us to be more alert on possible systematic uncertainties.  Hence
in the remainder of our paper we will only focus upon the five most clear cases:
Coma, A~1795, S\'ersic~159--03, MKW~3s and A~2052.

Is it possible to test which of the three models for the soft excess (power law,
thermal CIE model with cluster abundances, thermal CIE model with zero
metallicity) fits the data best?  Performing a formal F-test, only for the
outermost two annuli 7 \& 8 (radius between 6--12\arcmin) of the MKW~3s and
A~2052 clusters the thermal CIE model with cluster abundances is much better
than the power law model or the thermal CIE model with zero metallicity.  Hence,
in Sect.~\ref{sec:models} we investigate in more detail the properties and
physical implications for these three models in all five clusters with a
significant soft excess.  But first we consider the spectral signature of the
soft excess in these five clusters.

\subsection{Spectral signature of the soft excess:  \ion{O}{vii}
emission\label{sec:o7}}

\begin{figure}
\resizebox{\hsize}{!}{\includegraphics[angle=0]{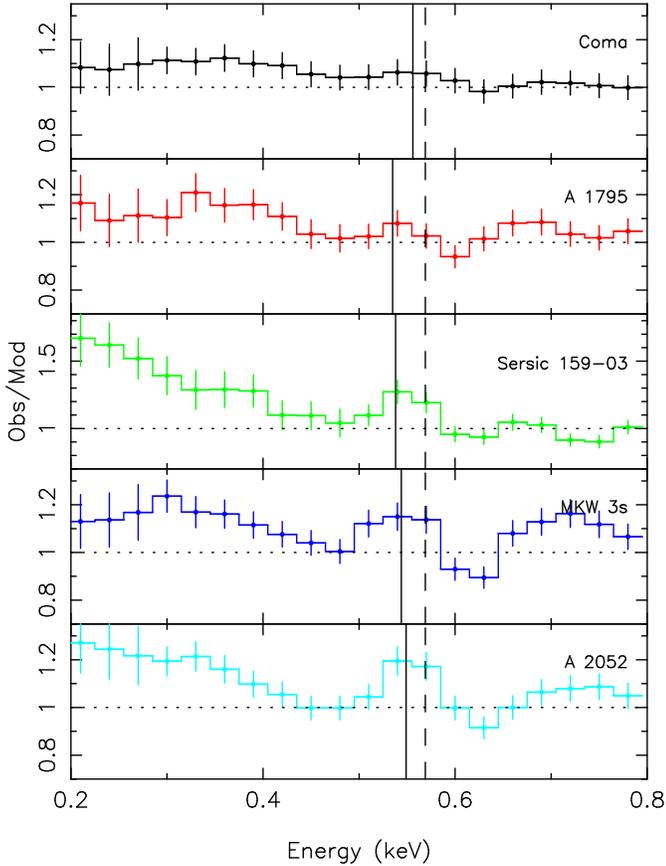}}
\caption{Fit residuals with respect to the two temperature model
for the outer 4--12\arcmin\ part of five clusters.
We have included the systematic background error in the fit,
but have excluded it in this plot. We indicate in each panel the position of
the \ion{O}{vii} triplet in the cluster restframe by a solid line and
in our Galaxy's rest frame by a dashed line at 0.569~keV (21.80~\AA).
The fit residuals for all instruments (MOS, pn) are combined.
The instrumental resolution at 0.5~keV is $\sim$60~eV (FWHM).
A similar
plot but now for the 0.5--4.0\arcmin\ range is shown as Fig.~\ref{fig:2tresin}.}
\label{fig:2tres}
\end{figure}

The fit residuals with respect to our basic two temperature model in the outer
part of the clusters (4\arcmin--12\arcmin) show in several cases apart from a
soft excess also clear evidence for an emission blend of \ion{O}{vii} around
0.54~keV.  This is demonstrated in Fig.~\ref{fig:2tres}.  The centroids of this
feature could be determined for S\'ersic~159--03, MKW~3s and A~2052
(Table~\ref{tab:o7centroid}).
\begin{table}[!h]
\caption{Centroid of the \ion{O}{vii} triplet (in eV) as measured
from the fit residuals in the 4--12\arcmin\ range, assuming it 
is emitted in our Galaxy
(no redshift, second column) or in the cluster (cosmological redshift,
third column). }
\label{tab:o7centroid}
\centerline{
\begin{tabular}{|lrr|}
\hline
Cluster & Galaxy & cluster   \\
\hline
S\'ersic~159--03  & 544$\pm$10 & 576$\pm$10 \\
MKW~3s            & 537$\pm$14 & 561$\pm$14 \\
A~2052            & 550$\pm$10 & 570$\pm$10 \\
\hline\noalign{\smallskip}
\end{tabular}
}
\end{table}

\begin{figure}
\resizebox{\hsize}{!}{\includegraphics[angle=0]{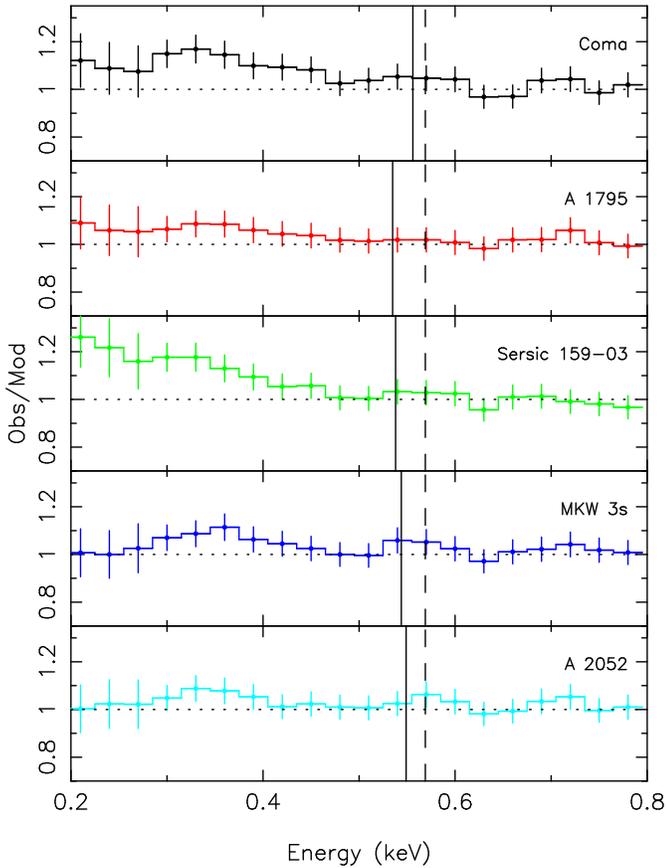}}
\caption{Fit residuals with respect to the two temperature model
for the inner 0.5--4.0\arcmin\ part of five clusters.
We have included the systematic background error in the fit,
but have excluded it in this plot. 
The fit residuals for all instruments (MOS, pn) are combined.
We indicate in each panel the position of
the \ion{O}{vii} triplet in the cluster restframe by a solid line and
in our Galaxy's rest frame by a dashed line at 0.569~keV (21.80~\AA).}
\label{fig:2tresin}
\end{figure}

Note that for these plots the systematic background errors were excluded because
of the following reasons.  In the inner parts (Fig.~\ref{fig:2tresin}) the
background can be neglected anyway (the modeling errors are dominated there by
the systematic effective area uncertainties and data Poisson noise).  In the
outer parts (Fig.~\ref{fig:2tres}) we will show below (Fig.~\ref{fig:fex}) that
the degree of soft X-ray excess is commensurate with the 1/4 keV sky background
enhancement in the vicinity of these clusters - an effect at the level of our
35~\% adopted systematic background uncertainty at low energies.  Thus the
question concerning whether the outer soft excess is a background rather than
cluster phenomenon is already clear, and has to be addressed by examining the
{\it original} spectral data of the excess.

The resonance, intercombination and forbidden line of \ion{O}{vii} have energies
of 574, 569 and 561~eV, respectively; for a low density plasma at 0.2~keV
temperature, the centroid of the triplet (including satellite lines) has an
energy 568.7~eV, and for the entire 0.05--3~keV temperature range this centroid
does not vary by more than 0.5~eV.

It is evident from Table~\ref{tab:o7centroid} that the measured centroid is in
agreement with an origin at the cluster redshift.  However, we cannot fully
exclude that (a part of) it is of Galactic origin.  This would be the case in
particular if the triplet emission would be dominated by the forbidden line
(561~eV), which is the case if photoionization of this medium is important.
Although photoionization of the interstellar medium by the diffuse Galactic
radiation field may occur, it is not obvious that it should be as strong as for
example in AGN environments.  Some caution is also needed, because the centroid
may be slightly shifted due to the influence of the \ion{O}{viii} Ly$\alpha$
line of the cluster at 653~eV, which is redshifted by about 30~eV towards any
Galactic \ion{O}{vii} triplet.

In order to investigate the origin of the \ion{O}{vii} emission further, we have
performed an additional set of spectral fits.  Contrary to our earlier analysis,
we took for the soft excess not a relatively cool plasma with cluster abundances
and cosmological redshift, but a plasma with solar abundances and zero redshift
(i.e.  a Galactic origin).  Restricting again to the 4--12\arcmin\ range,
$\chi^2$ increases (with respect to the hypothesis of a cluster origin) from 864
to 884, 671 to 743 and 798 to 948, for S\'ersic~159--03, MKW~3s and A~2052,
respectively.  The number of degrees of freedom in all cases is 702.  When we
perform a formal F-test, the Galactic origin can be rejected only marginally for
MKW~3s (at the 91~\% confidence level) and more significantly for A~2052 (at the
99~\% confidence level).  The $\chi^2$ increase for S\'ersic~159--03 is not
significant (formal level:  62~\%).

Finally we have checked whether uncertainties in the gain of the EPIC camera's
may affect our results.  The gain slowly degrades as a function of time, but in
principle the data processing takes this into account.  We have measured the
energy of the instrumental background line of Al-K at 1.487~keV for MKW~3s and
A~2052.  For both fields, the measured centroid for all three EPIC camera's
agrees within 1~\% with the predicted energy:  the measured average line energy
for MKW~3s is only 0.4~\% smaller than predicted, and for A~2052 0.8~\%.  For
the pn camera, the measured energy of the Cu-K instrumental line even agrees
within 0.1~\% with the predicted energy of 8.041~keV.  We conclude that there is
no evidence for significant energy scale uncertainties.

The \ion{O}{vii} emission line is difficult to recognize in the central parts of
the clusters.  This is illustrated in Fig.~\ref{fig:2tresin}.  This does not
imply that there is no such line emission in the central parts, but its signal
may simply be lost due to the strong intensity of the radiation from the hot
gas, which is in particular bright in the cluster core.

We conclude that in particular in the outer parts of some of our clusters there
is evidence for \ion{O}{vii} line emission, and that at least a significant
component of the emission resposible for this line feature is of cluster origin.

\section{Sky background variations\label{sec:sky}}

In order to test if the excess can be explained by sky variations of the diffuse
X-ray background, we have collected intensity maps of the 1/4~keV band of the
ROSAT PSPC (Snowden et al.  \cite{snowden97}; we took the data from NASA's
skyview facility).  Table~\ref{tab:pspc25} lists the count rates for all
clusters as well as the background field in three annular regions:  the
innermost region with a radius of 15\arcmin\ typically spans the XMM-Newton
field of view, the second and third annuli give the background in the annuli
between 1\degr--2\degr\ and 2\degr--5\degr\ from the cluster.  For the
background field, we have determined the exposure-averaged values for the 8
individual fields.  The variation among those fields is consistent with the
adopted $\sim$35~\% variation at low energies.  For the brightest clusters
(Coma, Virgo), the enhanced emission due to the cluster itself is clearly
visible in the innermost annulus.

\begin{table}[!h]
\caption{Average ROSAT PSPC sky intensity for different regions
in the 1/4~keV band, in units of $10^{-3}$~counts\,s$^{-1}$\,arcmin$^{-2}$.}
\label{tab:pspc25}
\centerline{
\begin{tabular}{|lrrr|}
\hline
Cluster & 0\arcmin--15\arcmin\ & 1\degr--2\degr\ & 2\degr--5\degr\ \\
\hline
Coma cluster          &  2.07&  1.21&  1.37 \\
A 1795                &  1.31&  1.07&  1.10 \\
A 4059                &  0.97&  0.86&  0.89 \\
S\'ersic~159--03      &  1.21&  0.99&  1.02 \\
Virgo                 &  2.44&  1.20&  1.44 \\
A 1835                &  1.03&  0.99&  1.01 \\
MKW 3s                &  1.28&  1.19&  1.20 \\
A 2052                &  1.36&  1.18&  1.21 \\
NGC 533               &  0.46&  0.47&  0.47 \\
A 496                 &  0.83&  0.86&  0.84 \\
A 1837                &  0.50&  0.58&  0.57 \\
Hydra A               &  0.64&  0.60&  0.61 \\
A 262                 &  0.53&  0.50&  0.50 \\
2A 0335+096           &  0.55&  0.52&  0.52 \\
\hline
Background            &  0.90&  0.86&  0.87 \\
\hline\noalign{\smallskip}
\end{tabular}
}
\end{table}

The spectrum of the diffuse background at high Galactic latitudes has been
modeled by Kuntz \& Snowden (\cite{kuntz}). It consists of four components.
We list them here with their relative contributions in the 0.2--0.3~keV band: 
1. the local hot bubble with temperature 0.11~keV (48~\%);
2. A soft distant component with temperature 0.09~keV (34~\%); 
3. A hard distant component with temperature 0.18~keV (7~\%);
4. The extragalactic power law with photon index 1.46 (11~\%).
In the soft 0.2--0.3~keV band, most of the diffuse sky background is therefore
due to local warm gas.

\begin{figure}
\resizebox{\hsize}{!}{\includegraphics[angle=-90]{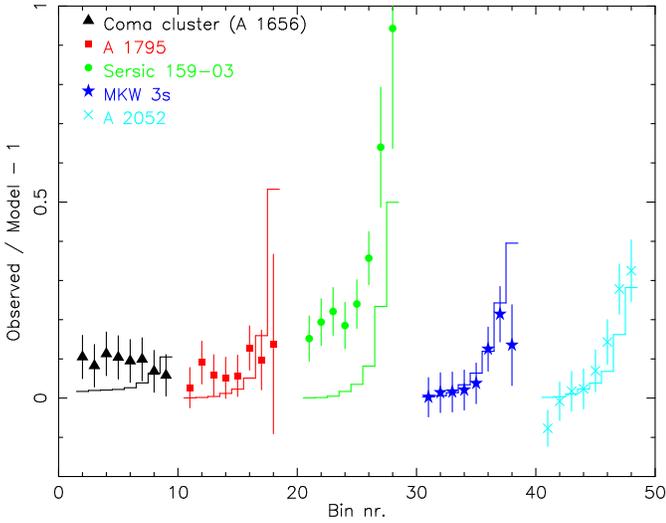}}
\caption{Soft excess in the 0.2--0.3~keV band as compared with a
two temperature model (data points with error bars). The error bars in this
plot are the statistical error bars plus a 5~\% of the source spectrum systematic
uncertainty; the 35~\% of the background uncertainty is {\it not} included.
The solid histogram represents the excess background near the cluster,
based upon the subtracted
background in the XMM-Newton data, multiplied by the excess percentages
derived from the 1/4~keV PSPC images. 
For the $k$th cluster the  values for annulus $j$ are plotted at bin
number $10(k-1)+j$.}
\label{fig:fex}
\end{figure}

All the five clusters with a soft excess appear to be located in regions with
enhanced soft X-ray background as compared with our background field, as is
evident from Table~\ref{tab:pspc25}.  The background in the 1--2\degr\ annulus
is enhanced by 41, 24, 15, 38 and 37~\% for Coma, A~1795, S\'ersic~159--03,
MKW~3s and A~2052, respectively.  In Fig.~\ref{fig:fex} we plot the fractional
soft excess in the 0.2--0.3~keV band, as compared with the two temperature
model.  The error bars in this plot do not contain the 35~\% systematic
background uncertainty that we took into account otherwise in our analysis.
Instead of that we also plot an estimate of the contribution to the measured
soft excess due to the local X-ray background enhancement.  This estimate is
given by the subtracted background in the XMM-Newton data, multiplied by the
excess percentages mentioned above as derived from the PSPC data.  This is
justified here since at low energies the total background in the EPIC camera's
is dominated by the diffuse X-ray background and the contribution of particles
is small.

\begin{figure}
\resizebox{\hsize}{!}{\includegraphics[angle=-90]{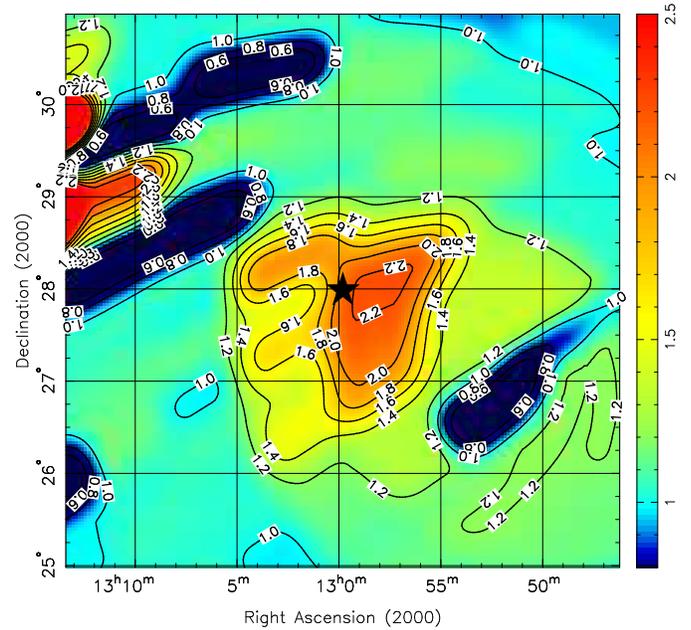}}
\caption{ROSAT PSPC 1/4~keV band image around the Coma cluster (star).
The original pixel size is about 40\arcmin, the image has been smoothed with
a Gaussian with a $\sigma$ of 10\arcmin. Units are 
$10^{-3}$~counts\,s$^{-1}$\,arcmin$^{-2}$.}
\label{fig:im_coma}
\end{figure}

\begin{figure}
\resizebox{\hsize}{!}{\includegraphics[angle=-90]{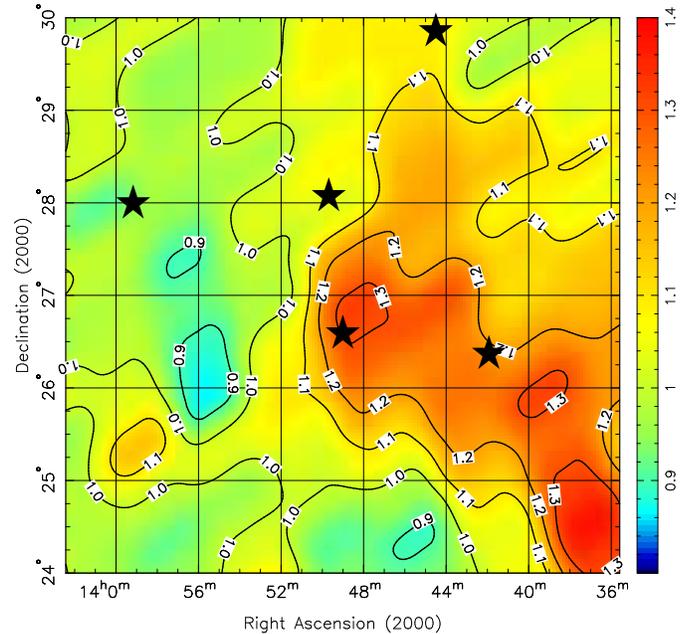}}
\caption{As Fig.~\ref{fig:im_coma}, but for A~1795. Members of the Bootes
supercluster are indicated by stars, from right to left: 
A~1775, A~1781, A~1795, A~1800 and A~1831.
}
\label{fig:im_a1795}
\end{figure}

\begin{figure}
\resizebox{\hsize}{!}{\includegraphics[angle=-90]{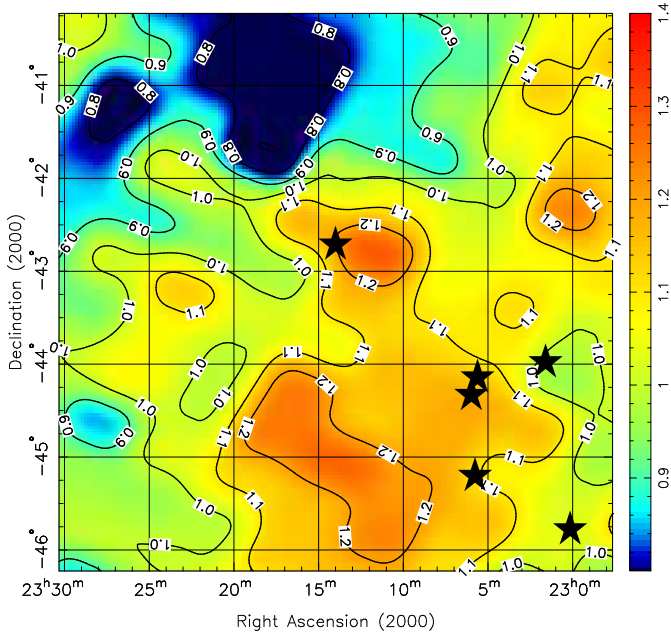}}
\caption{As Fig.~\ref{fig:im_coma}, but for S\'ersic~159--03.
Members of the supercluster
SCL~206 are indicated by stars, from top to bottom: 
S\'ersic~159--03, AS~1080, A~3969, A~3972, A~3970 and A~3952.}
\label{fig:im_as1101}
\end{figure}

\begin{figure}
\resizebox{\hsize}{!}{\includegraphics[angle=-90]{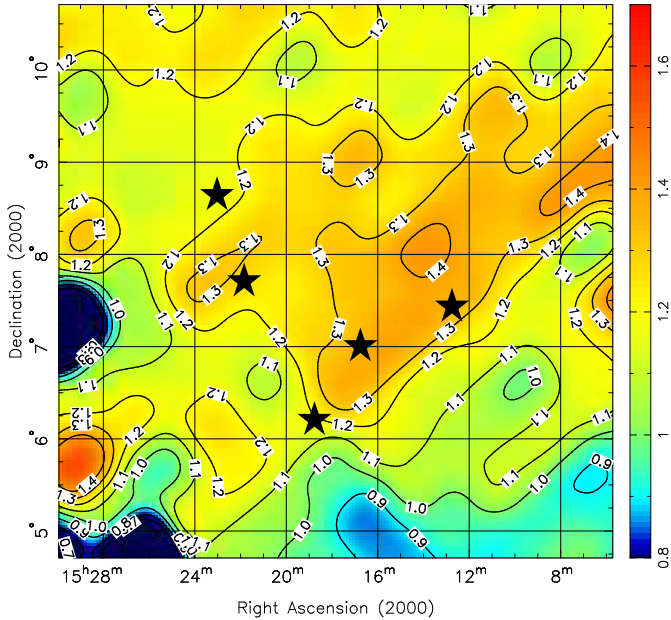}}
\caption{As Fig.~\ref{fig:im_coma}, but for the region around MKW~3s
and A~2052.
Members of the Hercules supercluster
are indicated by stars, from right to left:
A~2040, A~2052, A~2055, MKW~3s and A~2063. 
}
\label{fig:im_mkw3s}
\end{figure}

In A~1795, MKW~3s and A~2052, the observed soft excess is clearly consistent
with an enhancement of the diffuse soft X-ray sky background in the vicinity of
each cluster, as revealed earlier by ROSAT.  This is not the case for Coma, and
definitely not the case for S\'ersic~159--03, although in this last case the
increase of the soft excess beyond 4\arcmin\ (annuli 6--8) could be due to this
sky background enhancement.  Note also that all five clusters except Coma are
located not only in regions with enhanced soft X-ray emission, but interestingly
they are also close to the boundaries of these large scale structures (see
Figs.~\ref{fig:im_coma}--\ref{fig:im_mkw3s}).  We come back to this in
Sect.~\ref{sec:discussion}.

The enhancement of the diffuse soft X-ray sky background near at least three of
these clusters thus corresponds well with the detected soft X-ray excess.  Its
magnitude is consistent with the 35~\% systematic error applied to the lowest
energy part of the spectra.  The third, soft component added to our model
describes both shape and magnitude of this enhancement quite well, as is
evidenced by the improvement of the fit.

\section{Models for the soft excess\label{sec:models}}

We have seen before that it is not always possible to discriminate between the
three different spectral models, based upon goodness of fit criteria.  Therefore
we consider in this section the implications and derived physical parameters for
all three scenarios.

\subsection{Power law model}

A non-thermal interpretation of the soft excess was discussed in section 1.
More recently, McCarthy et al.  (\cite{mccarthy}) proposed that a power law
component contributing $\sim$30~\% of the total luminosity in the central parts
of several clusters provides a good fit to the ASCA/PSPC spectra of these
clusters.  They couple this to the cooling flow problem.  The relativistic
electrons may either be genuinely diffuse or associated with AGNs.  It should be
emphasized, however, that any widespread existence of such electrons in the
intracluster medium cannot easily be accounted for by processes other than
diffusive shock acceleration of cosmic rays, which {\it necessarily} implies the
intracluster presence of relativistic protons carrying at least ten times more
pressure than the electrons (see Lieu et al.  \cite{lieu1999b}).  Thus, while
for a central cluster region this pressure may even provide the means of halting
any cooling flows, for the outer regions a large cosmic ray pressure may pose
stability problems - as also alluded to in section 1, the cosmic rays cannot be
confined.

\begin{figure}
\resizebox{\hsize}{!}{\includegraphics[angle=-90]{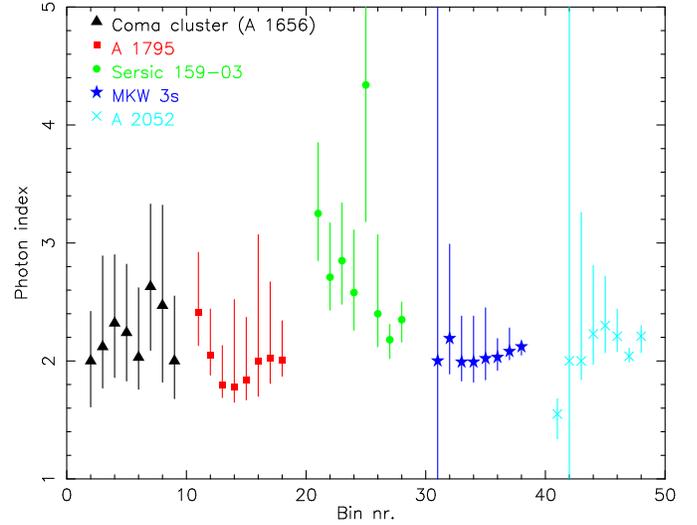}}
\caption{Best fit photon indices $\Gamma$ of the soft component
as a function of annulus. For the
$k$th cluster the $\Gamma$ values for annulus $j$ are plotted at bin
number $10(k-1)+j$.}
\label{fig:gam}
\end{figure}

\begin{figure}
\resizebox{\hsize}{!}{\includegraphics[angle=-90]{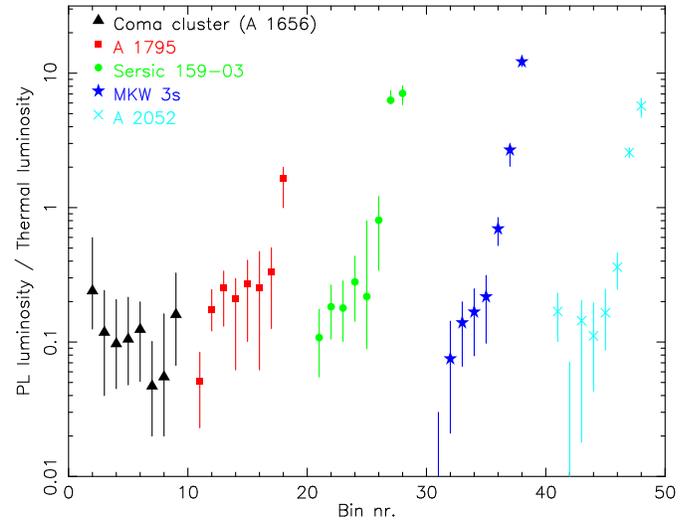}}
\caption{Luminosity of the soft component in the $0.2-10$~keV band
divided by the luminosity of the thermal cluster gas,
as a function of annulus. For the
$k$th cluster the  values for  annulus $j$ are plotted at bin
number $10(k-1)+j$.}
\label{fig:lum}
\end{figure}

The derived photon index of the power law component for the five clusters with a
significant soft excess is shown in Fig.~\ref{fig:gam}, the luminosity in the
full 0.2--10~keV band as a fraction of the luminosity of the thermal gas is
shown in Fig.~\ref{fig:lum}.

For Coma, the soft excess contributes about 10~\% to the total luminosity of all
annuli, while for the other clusters the relative contribution of the power law
increases from about 10~\% in the center to 100~\% or even a factor of 10 larger
in the outer parts.  It should be noted, however, that for the outermost annuli
of MKW~3s and A~2052, where the relative luminosity of the power law is very
large, a thermal model with cluster abundances for the soft excess yields a
better fit.

The photon index as a function of radius is constant for most clusters, at a
value close to 2.  Only in S\'ersic~159--03 the photon index decreases from
about 3 in the center to about 2.3 in the outer parts.

\begin{figure}
\resizebox{\hsize}{!}{\includegraphics[angle=-90]{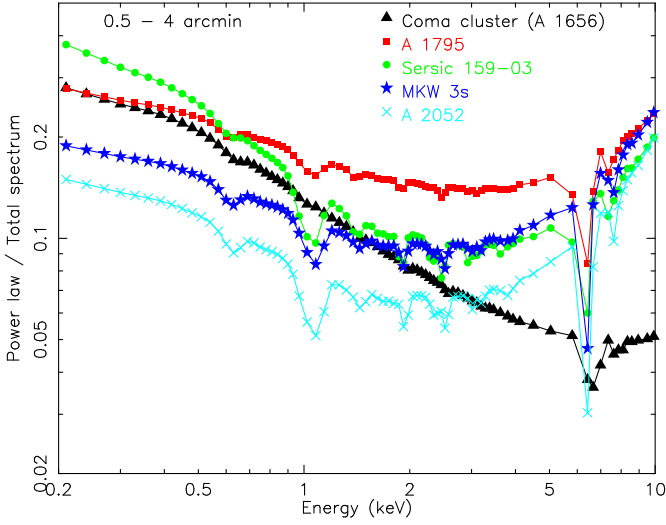}}
\caption{Fraction of the model spectrum caused by the soft power law
component, for the 0.5\arcmin\ -- 4\arcmin\ range. The dip near
6~keV is due to the Fe-K complex.
}
\label{fig:modpow1}
\end{figure}

\begin{figure}
\resizebox{\hsize}{!}{\includegraphics[angle=-90]{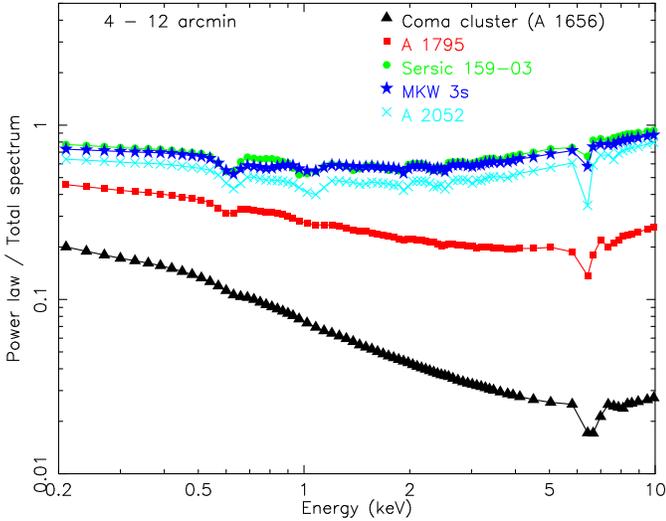}}
\caption{Fraction of the model spectrum caused by the soft power law
component, for the 4\arcmin\ -- 12\arcmin\ range.
}
\label{fig:modpow2}
\end{figure}

The luminosity-weighted photon indices as well as the total luminosity of the
soft excess integrated over all annuli are listed in Table~\ref{tab:pow}.  It is
evident that the power law component contains a significant fraction of the
total luminosity of the cluster, in particular at large radii.  This is
illustrated in Figs.~\ref{fig:modpow1} and \ref{fig:modpow2}, where we show the
fraction of the total model spectrum caused by the soft power law component, for
the inner parts of the cluster and the outer parts.

\begin{table}[!h]
\caption{Power law fits to the soft excess. Luminosities
of the power law ($L_{\mathrm{PL}}$) and thermal plasma $L_{\mathrm{th}}$
are in the 0.2--10~keV band in the rest frame of the cluster.}
\label{tab:pow}
\centerline{
\begin{tabular}{|lrrr|}
\hline
Cluster & $\Gamma$ & $L_{\mathrm{PL}}$ & $L_{\mathrm{th}}$ \\
        &   & ($10^{37}$~W) & ($10^{37}$~W) \\
\hline
Coma              & 2.2$^{+0.6}_{-0.4}$ & 1.0$\pm$0.3 & 10.5 \\
A 1795            & 1.9$^{+0.6}_{-0.2}$ & 4.8$\pm$0.9 & 19.9 \\
S\'ersic~159--03  & 2.6$^{+0.5}_{-0.4}$ & 1.3$\pm$0.2 & 4.5 \\
MKW 3s            & 2.1$^{+0.3}_{-0.2}$ & 1.7$\pm$0.2 & 4.0 \\
A 2052            & 2.1$^{+0.4}_{-0.2}$ & 1.2$\pm$0.1 & 3.0 \\
\hline\noalign{\smallskip}
\end{tabular}
}
\end{table}
\begin{figure}
\resizebox{\hsize}{!}{\includegraphics[angle=-90]{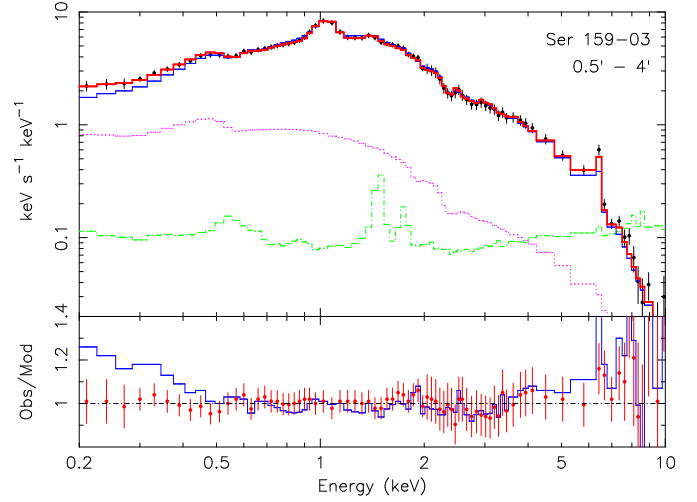}}
\caption{Spectral fits to the spectrum of S\'ersic~159--03
in the 0.5\arcmin\ -- 4\arcmin\ range. The data for MOS1, MOS2
and pn have been combined. {\it{Upper panel (spectra):}}
Data points with error bars: measured
background-subtracted spectrum (multiplied with energy for display
purposes); dash-dotted line: subtracted background; thin solid line:
best fit two temperature model with Galactic absorption; thick solid
line: best fit two temperature plus power law model; dotted line:
contribution of the power law component to the 
two temperature plus power law model fit. {\it{Lower panel (fit residuals):}}
thin solid line: two temperature model with Galactic absorption;
data points with error bars: two temperature plus power law model.
}
\label{fig:as1101in}
\end{figure}

Finally, as a typical example of the quality of the spectral fits, we show in
Fig.~\ref{fig:as1101in} the fit in the 0.5\arcmin--4.0\arcmin\ range of
S\'ersic~159--03.  It is seen that when the power law component is added, the
fit is in particular improved at low energies and at the Fe-K complex near
6.7~keV.

\subsection{Thermal emission from warm cluster plasma}

\begin{figure}
\resizebox{\hsize}{!}{\includegraphics[angle=-90]{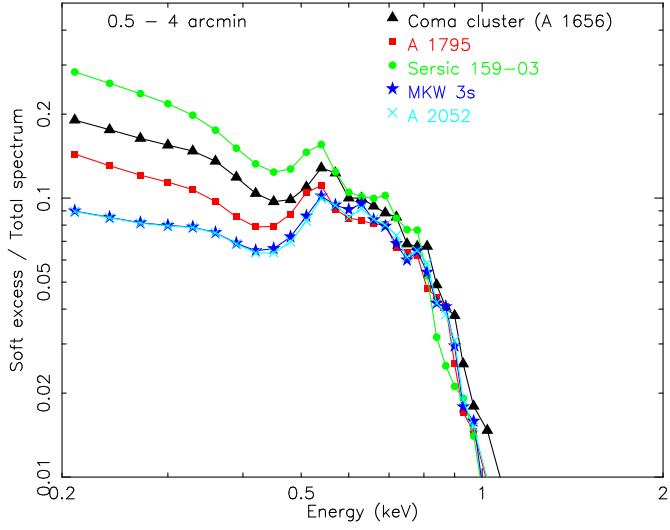}}
\caption{Fraction of the model spectrum caused by the soft thermal
component, for the 0.5\arcmin\ -- 4\arcmin\ range.
}
\label{fig:modcie1}
\end{figure}

\begin{figure}
\resizebox{\hsize}{!}{\includegraphics[angle=-90]{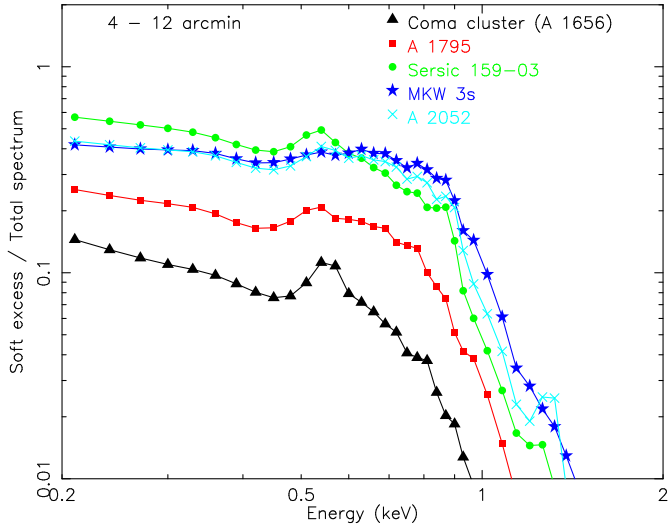}}
\caption{Fraction of the model spectrum caused by the soft thermal
component, for the 4\arcmin\ -- 12\arcmin\ range.
}
\label{fig:modcie2}
\end{figure}

An alternative interpretation of the soft excess is that it originates from a
thermal warm plasma accounting for a fair fraction of the missing baryons in the
present epoch, with temperature distinctly lower than that of the virialized
intracluster medium, as discussed in section 1.  We tested this possibility by
adding a third, thermal emission component to the original model, with the
abundances coupled to the abundances of the hotter cluster plasma.  In the outer
annuli of MKW~3s and A~2052 this provides a statistically better fit as compared
with a power law soft excess.  In this case, the soft component only contributes
with the spectrum below $\sim$\,1~keV (Figs.~\ref{fig:modcie1} and
\ref{fig:modcie2}), contrary to the power law model that dominates the spectrum
at large radii for almost all energies (Figs.~\ref{fig:modpow1} and
\ref{fig:modpow2}).

In all the five clusters, the best-fit temperature does not vary significantly
as a function of radius.  The average values for each cluster are given in
Table~\ref{tab:cie}.  In all these clusters, the temperature of the soft excess
is close to 0.2~keV.
\begin{table}[!ht]
\caption{Thermal plasma model fits to the soft excess. 
$Y_{\mathrm{cool}}$ represents the emission measure of the soft excess
(with temperature k$T$ as listed in the first column), while
$Y_{\mathrm{hot}}$ is the emission measure of the two hot components
added together. }
\label{tab:cie}
\centerline{
\begin{tabular}{|lrrr|}
\hline
Cluster & k$T$  & $Y_{\mathrm{cool}}$ & $Y_{\mathrm{hot}}$ \\ 
        & (keV) & ($10^{72}$~m$^{-3}$)& ($10^{72}$~m$^{-3}$) \\
\hline
Coma       &  0.187$\pm$0.011 &  9$^{+4}_{-2}$     &  63 \\ 
A 1795     &  0.200$\pm$0.008 & 17$^{+6}_{-2}$     & 138 \\
S\'ersic~159--03 &  0.180$\pm$0.006 & 16$^{+4}_{-2}$     &  42 \\
MKW 3s     &  0.230$\pm$0.012 &7.8$^{+1.0}_{-0.8}$ &  36 \\
A 2052     &  0.208$\pm$0.007 &6.9$^{+1.1}_{-0.7}$ &  27 \\
\hline\noalign{\smallskip}
\end{tabular}
}
\end{table}

\begin{figure}
\resizebox{\hsize}{!}{\includegraphics[angle=-90]{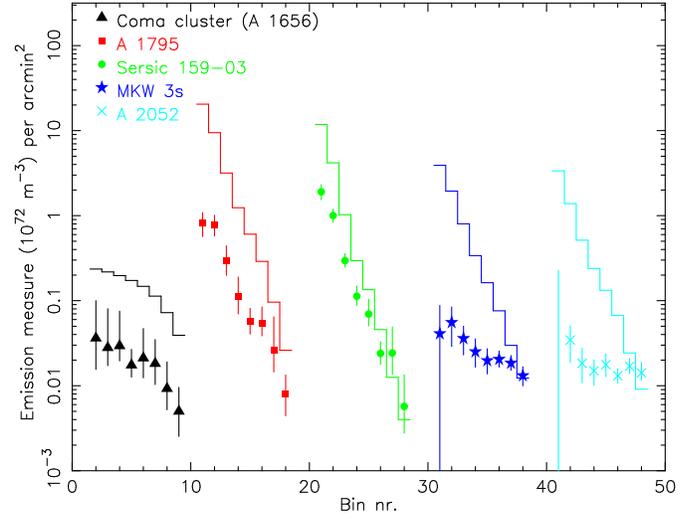}}
\caption{Emission measure per solid angle
of the soft component (data points with error bars) and of the hot
gas (both components added together, solid histogram),
as a function of annulus. The soft component is modeled here as warm
gas with cluster abundances. For the
$k$th cluster the values for  annulus $j$ are plotted at bin
number $10(k-1)+j$.}
\label{fig:temp}
\end{figure}

The radial distribution of the emission measure of the soft excess is shown in
Fig.~\ref{fig:temp}, together with the emission measure of the hot gas (both
temperature components of the hot component added together).  The data are
plotted as a surface brightness.  For Coma, the emission measure of the soft
excess is almost a constant fraction of the emission measure of the hot gas,
while for A~1795 and S\'ersic~159--03 it increases slowly with radius as
compared with the hot gas.  The same holds for MKW~3s and A~2052, but in both
cases the surface brightness of the soft component is more or less constant
within the error bars.
\begin{figure}
\resizebox{\hsize}{!}{\includegraphics[angle=-90]{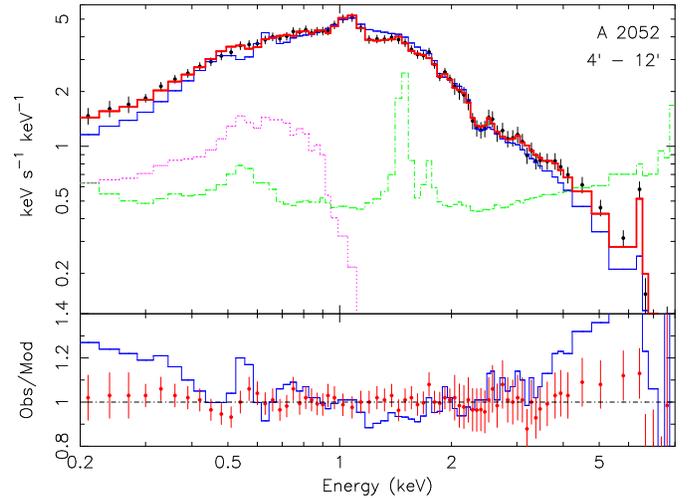}}
\caption{Spectral fits to the spectrum of A~2052
in the 4\arcmin\ -- 12\arcmin\ range. The data for MOS1, MOS2
and pn have been combined. {\it{Upper panel (spectra):}}
Data points with error bars: measured
background-subtracted spectrum (multiplied with energy for display
purposes); dash-dotted line: subtracted background; thin solid line:
best fit two temperature model with Galactic absorption; thick solid
line: best fit three temperature model; dotted line:
contribution of the warm thermal component to the 
three temperature  model fit. {\it{Lower panel (fit residuals):}}
thin solid line: two temperature model with Galactic absorption;
data points with error bars: best fit three temperature model.
}
\label{fig:a2052out}
\end{figure}

Again, as a typical example of the quality of the spectral fits, we show in
Fig.~\ref{fig:a2052out} the fit in the 4\arcmin--12\arcmin\ range of A~2052.  It
is seen that when the third, warm thermal component is added, the fit improves
for several energies, including the low- and high energy tail, but also for the
region between 0.5--2~keV.

\subsection{Thermal emission from warm gas with zero metallicity}

\begin{figure}
\resizebox{\hsize}{!}{\includegraphics[angle=-90]{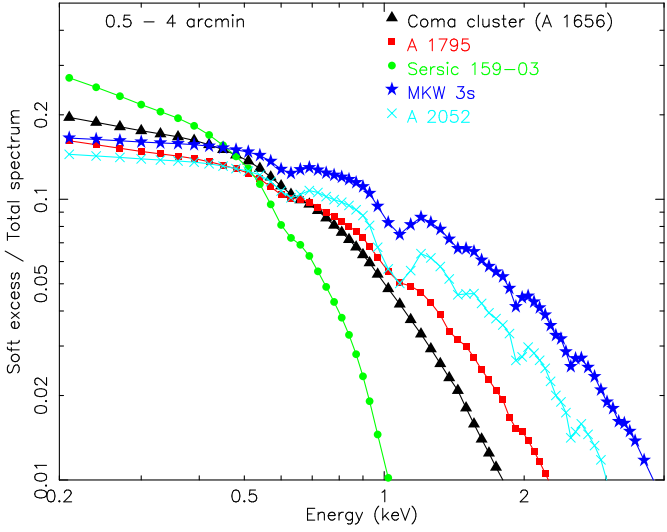}}
\caption{Fraction of the model spectrum caused by the soft thermal
component with zero metallicity, for the 0.5\arcmin\ -- 4\arcmin\ range.
}
\label{fig:modnul1}
\end{figure}

\begin{figure}
\resizebox{\hsize}{!}{\includegraphics[angle=-90]{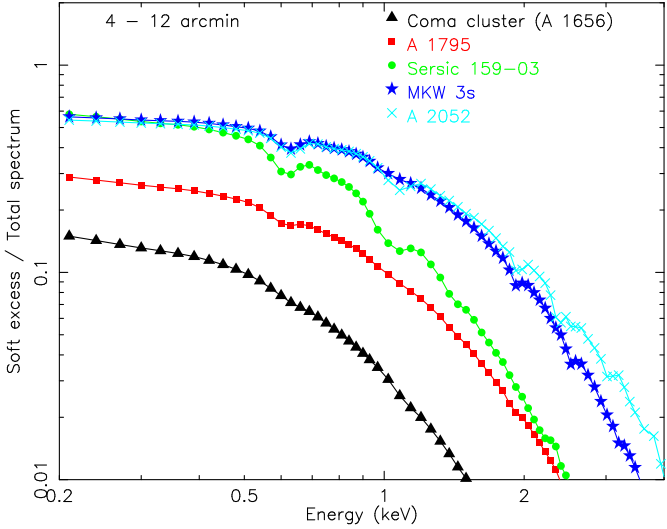}}
\caption{Fraction of the model spectrum caused by the soft thermal
component with zero metallicity, for the 4\arcmin\ -- 12\arcmin\ range.
}
\label{fig:modnul2}
\end{figure}

Instead of a thermal plasma with cluster abundances, we consider here a model
for the soft excess consisting of warm gas with zero metallicity (a pure H/He
mixture).  This could apply if for example the soft excess is caused by fresh,
not enriched material falling into the cluster.  Now the soft component
contributes to a larger range of the spectrum as compared with the model with
cluster abundances:  significant excess emission is produced below $\sim$\,2~keV
(Figs.~\ref{fig:modnul1} and \ref{fig:modnul2}).

Again, in all the five clusters, the best-fit temperature does not vary
significantly as a function of radius.  The average values for each cluster are
given in Table~\ref{tab:cienul}.  The temperatures for most clusters except
S\'ersic~159--03 are larger than in the case with cluster abundances for the
soft excess.  Also the emission measures are larger.

\begin{table}[!h]
\caption{Thermal plasma model fits to the soft excess, with zero
metallicity for the excess. 
$Y_{\mathrm{cool}}$ represents the emission measure of the soft excess
(with temperature k$T$ as listed in the first column), while
$Y_{\mathrm{hot}}$ is the emission measure of the two hot components
added together. }
\label{tab:cienul}
\centerline{
\begin{tabular}{|lrrr|}
\hline
Cluster & k$T$  & $Y_{\mathrm{cool}}$ & $Y_{\mathrm{hot}}$ \\ 
        & (keV) & ($10^{72}$~m$^{-3}$) & ($10^{72}$~m$^{-3}$) \\
\hline
Coma       &  0.50$\pm$0.09  &11$^{+2}_{-1}$ & 63 \\ 
A 1795     &  0.69$\pm$0.12  &31$^{+3}_{-2}$ & 138 \\
S\'ersic~159--03 &  0.19$\pm$0.04  &39$^{+11}_{-5}$ & 42 \\
MKW 3s     &  0.73$\pm$0.04  &17.1$^{+1.0}_{-0.9}$ & 36 \\
A 2052     &  0.61$\pm$0.05  &12.9$^{+5.4}_{-0.6}$ & 27\\
\hline\noalign{\smallskip}
\end{tabular}
}
\end{table}
\begin{figure}
\resizebox{\hsize}{!}{\includegraphics[angle=-90]{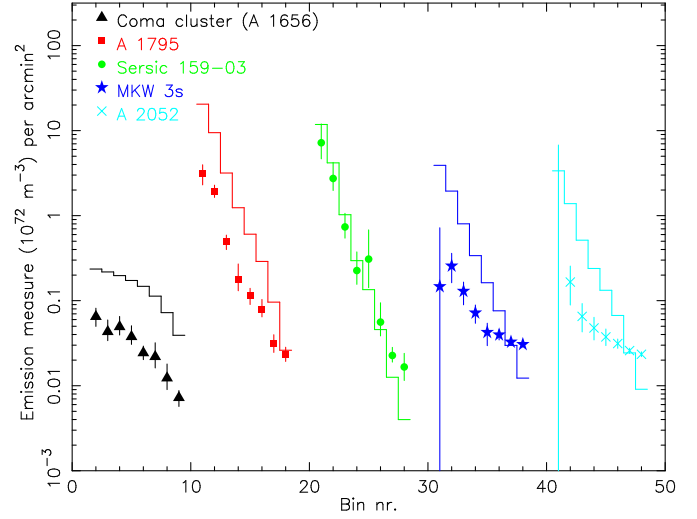}}
\caption{Emission measure per solid angle
of the soft component (modeled here as a thermal plasma with
zero metallicity; data points with error bars) and of the hot
gas (both components added together, solid histogram),
as a function of annulus. For the
$k$th cluster the values for  annulus $j$ are plotted at bin
number $10(k-1)+j$.}
\label{fig:tempnul}
\end{figure}

The radial distribution of the emission measure of the soft excess is shown in
Fig.~\ref{fig:tempnul}, similar to Fig.~\ref{fig:temp} for the case of a soft
excess with cluster abundances.  Qualitatively the same remarks can be made as
for that model (see previous section).  However, contrary to the model with
cluster abundances, the present model fails to explain the strong \ion{O}{vii}
line emission near 0.57~keV.

\section{Discussion\label{sec:discussion}}

We have studied here the presence and properties of a soft X-ray excess in a
sample of 14 clusters of galaxies.  Four of these clusters appear to have larger
absorption column densities than the column densities derived from 21~cm
observations.  In the remaining sample of 10 clusters, four show no direct proof
of significant excess emission.  The assessment of this absence is mainly
limited by the remaining small calibration uncertainties in the instrument
response.  One of the clusters (A~1835) is very compact, with a strong cooling
flow, making it hard to constrain a true soft excess.

In five clusters we have found evidence for a significant soft excess, stronger
than the uncertainties in calibration or background subtraction.  We discuss
those cases in the next section.  Here we conclude with a remark on our multi
temperature modeling.  Almost all of the studied clusters are cooling flow
systems where cooler thermal emission is expected in addition to the emission
from hotter gas, particularly for the central regions.  Based both upon our
modeling of RGS and EPIC spectra of several clusters (see
Sect.~\ref{sec:basicmodel}), as well as extensive spectral simulations done by
us, it is noted that because cooling flow does not appear to continue below
temperatures $\sim$ 1/3 that of the hot virial plasma and does not produce any
significant \ion{O}{vii} line emission, in our attempt to model cluster cores it
is sufficient to treat the emission measures of both components as free
parameters under the only constraint of $T_{\mathrm{cool}}=0.5T_{\mathrm{hot}}$.
This lower bound temperature for the cooling flow emission (typically between
one and a few keV) is also much higher than the characteristic temperature
(0.2~keV) of the soft excess.

\subsection{MKW~3s and A~2052}

Both clusters are close to each other on the sky (1.5\degr\ distance) as well as
in redshift (MKW~3s:  0.0455; A~2052:  0.0356), at a distance of about 60~Mpc.
Apparently they are located in a region of the sky with enhanced soft X-ray
emission distributed over an area of several degrees width, as is evidenced by
the PSPC maps (Fig.~\ref{fig:im_mkw3s}; Table~\ref{tab:pspc25}).  The flux of
the soft component as measured with ROSAT matches the flux of the soft excess as
measured here with XMM-Newton (Fig.~\ref{fig:fex}).  Our spectral fits, in
particular in the outer parts of the cluster, where the soft excess is
relatively strong, indicated that a thermal plasma model with cluster
metallicity provides a better spectral fit than a power law component.  The
temperature of the soft component of about 0.2~keV matches the temperature of
the "hard distant component" in the four component decomposition of the high
Galactic latitude modeling by Kuntz \& Snowden (\cite{kuntz}) mentioned earlier.
This hard component -- averaged over the sky -- contributes about 7~\% to the
total 0.2--0.3~keV photon number flux.  Since the diffuse soft X-ray emission in
this region of the sky is about 37~\% larger than on average, we conclude that
in this region this hard distant component is apparently 6 times brighter than
on average.

Kuntz \& Snowden (\cite{kuntz}) attribute this component in general to a
Galactic corona.  However, in a later paper (Kuntz et al.  \cite{kuntz2}) it is
shown that the PSPC data are not inconsistent with the hypothesis that a major
fraction of this hard component originates from the Warm-Hot Intergalactic
Medium (WHIM).  The PSPC data simply lack sufficient spectral resolution to
discriminate between both possible origins for this component.  Therefore we
conclude that at least a part of it may be due to localized diffuse emission
from the (supercluster) region.  This extragalactic origin is also indicated by
the presence of redshifted \ion{O}{vii} emission (Sect.~\ref{sec:o7}).

Is it possible that a large part of the enhanced 1/4~keV enhancement in this
region of the sky is caused by emission from filaments between clusters?  In
order to answer this question, we have investigated the distribution of galaxies
and clusters in this region of the sky.  Galaxies were taken from the Updated
Zwicky Catalog (Falco et al.  \cite{falco}); this catalog is an almost complete
catalog of galaxies up to a limiting magnitude of 15.5.  In the present field,
the redshift distribution is approximately bimodal, with two strong peaks near
redshifts of 0.035 (the redshift of A~2052) and 0.045 (the redshift of MKW~3s).
There are very few foreground or background galaxies catalogued in this region.
Galaxies of both redshift groups are found around the location of both A~2052
and MKW~3s, so that it looks as if two layers of galaxies are superimposed.
Both clusters are part of the southern tip of the Hercules supercluster, which
contains at least 12 Abell clusters and stretches NE-wards over a distance of at
least 30\degr\ (Einasto et al.  \cite{einasto}).  In Fig.~\ref{fig:im_mkw3s} we
also plot the locations of the Abell clusters of this Hercules supercluster in
the area surrounding MKW~3s and A~2052.  The local enhancement of the 1/4~keV
background near these two clusters coincides with this subconcentration of Abell
clusters, which further strengthens the case for an extragalactic origin of the
measured soft excess in these two clusters.

It is interesting to note that Bonemente et al.  (\cite{bonamente2002}) found
soft excesses using PSPC data in several clusters after subtracting a background
measured in a region outside the main cluster emission, yet a region which is
now found to be within the WHIM emitting radius.  The presence of a soft excess
in the data of Bonemente et al.  can be explained by an enhancement of the WHIM
towards the cluster as compared with the background region of Bonamente et al.
(due to enhanced filament density towards the connecting node of the filaments,
i.e.  the cluster).

Finally, we note that beyond the radius where the soft excess in both clusters
becomes significant (at about 4\arcmin\ radius or 300~kpc distance), the ambient
temperature of the hot cluster gas also starts decreasing significantly.  This
temperature drop is similar to what is observed in S\'ersic~159--03 (Kaastra et
al.  \cite{kaastra2001} and next section).

\subsection{S\'ersic~159--03}

A soft excess has been reported for this cluster previously by Bonamente et al.
(\cite{bonamente2001}), based upon ROSAT PSPC data.

As we have shown in the previous section, this cluster also shows evidence for
\ion{O}{vii} emission from the soft excess component, probably from
extragalactic origin.  A small supercluster (SCL~206) consisting of A~3952,
A~3969, A~3970 and A~3972 (Einasto et al.  \cite{einasto}) is found about
2\degr\ to the SW of the cluster, at a similar or slightly larger redshift than
S\'ersic~159--03.  This region shows also enhanced soft X-ray emission
(Fig.~\ref{fig:im_as1101}), but the enhanced region is much larger than this
supercluster.

In A~2052 and MKW~3s the soft excess has approximately a constant surface
brightness (Fig.~\ref{fig:temp}), and its flux is consistent with the PSPC
1/4~keV brightness in the region (Fig.~\ref{fig:fex}).  S\'ersic~159--03 has a
similar component as is evidenced by the relative increase of the soft excess at
large radii, consistent with the PSPC brightness (Fig.~\ref{fig:fex}).  However,
in addition there is a component with a $\sim$15~\% contribution in the central
part.  This component decreases in surface brightness for larger radii, more or
less similar to the brightness decrease of the hot gas (Fig.~\ref{fig:temp}).

The increase of the soft excess in S\'ersic~159--03 at large radii can thus be
ascribed to the same large scale soft X-ray component as in the other two
clusters, but the central enhancement cannot be explained this way.  In fact,
the fit residuals for the two temperature model show no significant \ion{O}{vii}
feature in S\'ersic~159--03 within a radius of 4\arcmin, but the low energy soft
excess continuum is present and strong (Fig.~\ref{fig:2tresin}).  In fact, a
power law model for the soft excess gives a somewhat better fit than a thermal
spectrum for these innermost radii.

We will now focus our attention on the soft excess in the inner parts of this
cluster.

\subsubsection{Warm intracluster gas}

We consider first the case that the warm gas coexists with the hot gas inside
the cluster.  As is shown in Fig.~\ref{fig:temp}, the emission measure of the
warm gas is typically a quarter of the emission measure of the hot gas.  We
assume here pressure equilibrium between both media.  For a central temperature
of 2.4~keV and a central hydrogen density of $2\times 10^4$~m$^{-3}$ (Kaastra et
al.  \cite{kaastra2001}), the warm gas at 0.2~keV temperature must have an
overdensity of a factor of $\sim$10 and a spatial filling factor $f$ of 0.2~\%.
Such dense, relatively cool gas would however cool down very rapidly, on a time
scale of $10^5$~year.  On the other hand, if the warm gas is contained in
magnetic pressure dominated loops that are in pressure equilibrium with the hot
gas, the gas pressure in the warm loops may be much smaller, resulting in a
lower gas density and hence a longer cooling time.  Such magnetic loops could be
similar to the loops advocated by Norman \& Meiksin (\cite{norman}).  In their
model, magnetic reconnection of cooler loops with hotter loops that are coupled
to the hot intracluster gas causes effectively the cooling rate to decrease by
an order of magnitude.  A strongly reduced cooling rate is indeed observed in
S\'ersic~159--03 (Kaastra et al.  \cite{kaastra2001}) and almost all our cooling
flow clusters (Peterson et al.  \cite{peterson2002}).  However, in these
clusters the amount of gas with temperatures below a few times the hot gas
temperature is very small, and it is hard to imagine why there should be little
cooling gas present at temperatures of $\sim$0.8~keV, but large amounts around
$\sim$0.2~keV.

\subsubsection{Nonthermal radiation due to magnetic reconnection}

Another possibility is of course that the soft excess is due to nonthermal
radiation caused by the reconnection of such magnetized loops.  In the central
4\arcmin\ of S\'ersic~159--03 the 0.2--10~keV luminosity of the soft excess is
about a quarter of the luminosity of the cooling gas (the coolest of the two hot
temperature components in our spectral fit).  Therefore the acceleration
mechanism causing the putative nonthermal emission will have to be efficient.  A
full investigation of the implications for a reconnection model is out of the
scope of this paper, however.

\subsubsection{Inverse Compton scattering}

Inverse Compton scattering as a possible mechanism for the production of a soft
excess was discussed in section~1.  However, in the case of S\'ersic~159--03
this model leads to a very large pressure of the cosmic ray electrons, as
already demonstrated by Bonamente et al.  (\cite{bonamente2001}).  The pressure
can be several times the gas pressure.

\subsubsection{Warm intercluster gas}

Another possible explanation for the central soft excess is that it is due to
thermal emission from warm gas that is located outside the cluster, for example
in the form of warm filaments seen in projection in front of or behind the
cluster.  The decreasing surface brightness of the soft excess emission then
implies that these filaments are probably oriented along the line of sight,
towards the direction to the cluster center.  If this warm gas is associated
with structures in the galaxy distribution, then we should see a large amount of
foreground or background galaxies superimposed on S\'ersic~159--03.
Unfortunately the available redshift data near this cluster region are very
sparse:  in fact, the cD galaxy of S\'ersic~159--03, ESO 291--9, is the only
galaxy with a measured redshift within 16\arcmin\ from the cluster center.

The emission measure of the soft excess is $1.6\times 10^{73}$~m$^{-3}$
(Table~\ref{tab:cie}).  For a cylindrical filament with a radius $r$ of
4\arcmin\ (400~kpc) and a density $n$ of 1000~m$^{-3}$ (the hot gas density in
S\'ersic~159--03 at a radius of 4\arcmin, cf.  Kaastra et al.
\cite{kaastra2001}), seen in projection in front of the cluster, the implied
column length $L$ is 1~Mpc.  It scales proportional to $(nr)^{-2}$.  These
dimensions are representative of warm filaments (e.g.  Cen \& Ostriker
\cite{cen1999}).  Previous detections of similar filamentary structures with Mpc
sizes were based upon correlations between low surface brightness X-ray
structures as seen with the PSPC and galaxy count overdensities (Wang et al.
\cite{wang}; Kull \& B\"ohringer (\cite{kull}; Scharf et al.  \cite{scharf}).

It is quite natural to expect that the filament density increases towards the
nodes (clusters); since the X-ray emissivity scales proportional to density
squared, the soft X-ray excess will be biased towards the densest regions close
to the cluster; this is for example contrary to absorption line studies of these
filaments, where the scaling is proportional to density (not squared).  Given
the uncertain density and temperature structure of the filaments (at least as
can be derived from the present data) we do not attempt to derive detailed mass
profiles here.

\subsection{A~1795}

A~1795 is a member of the Bootes supercluster (Einasto et al.  \cite{einasto}).
The region shown in Fig.~\ref{fig:im_a1795} forms the Western boundary of this
supercluster, which has a diameter of about 11\degr.  A~1795 is the X-ray
brightest cluster of this supercluster, its direct neighbours A~1775 and A~1800
are a factor of 5--6 fainter, while all other supercluster members are even
fainter than these last two clusters.

In A~1795, most of the soft excess emission, in particular at large radii, is
consistent with the enhanced soft X-ray background found with the ROSAT PSPC
(Figs.~\ref{fig:fex} and \ref{fig:im_a1795}).  The location of A~1795 near the
boundary of a region of enhanced soft X-ray emission, as well as the indication
of a weak redshifted \ion{O}{vii} line are suggestive for a similar origin of
warm intercluster gas as found in MKW~3s, A~2052 and Coma.  However, more
evidence is needed to prove the case for A~1795.  Remarkably, the soft excess as
measured with EUVE (Mittaz et al.  \cite{mittaz98}) is much stronger than we
find here using our XMM-Newton data.  This may indicate a very soft X-ray
spectrum, since EUVE has more sensitivity in the EUV band as compared with
XMM-Newton.

\subsection{Coma}

The Coma cluster is one of the first clusters where a soft excess has been
reported (Lieu et al.  \cite{lieucoma}).  In our present study of the central
part of Coma we find a soft X-ray excess of about 10~\%, almost constant in
surface brightness.  Most of the soft X-ray emission in the Coma cluster is
likely to be due to the cluster itself, as Fig.~\ref{fig:im_coma} shows that the
local soft X-ray enhancement is centered around Coma.  Note that Bonamente et
al.  (\cite{bonamente2002}) using PSPC data obtained a $\sim$~20~\% excess in
the overlapping region.  The discrepancy can be understood because of the
different spectral sensitivity as a function of energy for the PSPC as compared
to EPIC; also, at low energies the spectral redistribution function of both
instruments has strong low energy tails.  Therefore comparing fractional
excesses in count rates from different instruments is not very helpful.

Our spectral fits are not able to discriminate between the different spectral
models for the soft excess.  However, there is a big difference between the
models with a power law and warm thermal cluster gas.  For the power law model,
the derived oxygen abundance of the cluster is about 0.48 times solar, while for
the warm gas scenario the oxygen abundance is 0.07 times solar.  In both cases,
the best fit iron abundance is about 0.26--0.29 times solar.  The low oxygen
abundance in the warm gas scenario is driven by the absence of strong
\ion{O}{vii} emission in the central part of the Coma cluster, in combination
with the excess continuum emission below 0.6~keV.  Unfortunately it is not well
possible to derive firm conclusions on the preferred soft excess model from the
measured oxygen abundance.  RGS spectra of most of the clusters of our sample
yield oxygen abundances centered around 0.2--0.4 times solar (Tamura et al.
\cite{tamura2002}).  This is just between the values implied for both scenarios.

A power law component in Coma was found before using BeppoSAX PDS data
(Fusco-Femiano et al.  \cite{fusco}).  Their best fit model yields a photon
index of 1.8, with a confidence range of 0.7--3.6.  The corresponding 20--80~keV
flux was $2.2\times 10^{-14}$~W\,m$^{-2}$.  In the power law scenario, our best
fit model ($\Gamma = 2.2^{+0.6}_{-0.4}$) predicts a 20--80~keV flux of $7\times
10^{-15}$~W\,m$^{-2}$, a factor of three below the value found by Fusco-Femiano
et al.  However, if we put $\Gamma$ at our lower limit of 1.8, we get a hard
X-ray flux that matches exactly the BeppoSAX measurements.  It should be noted
further that the error bars on the photon index of the BeppoSAX spectrum are
rather large, and that our current extraction region (radius less than
12\arcmin) does not cover the full Coma cluster, so that in our case we might
miss a part of the power law flux.

Also, our modeling of a power law in the central parts of Coma are in good
agreement with previous EUVE data.  We find a power law contribution of
20--30~\% percent at 0.2~keV when the power law is included in the model (see
Figs.~\ref{fig:modpow1} and \ref{fig:modpow2}), which is in reasonable agreement
with the findings of Bowyer et al.  (\cite{bowyer1999}).

Finally, Cheng (\cite{cheng}) has argued that about 6~\% of the soft excess in
the central parts of Coma could be due to emission from individual elliptical
galaxies in the cluster.  A natural test of this model would be to correlate the
soft X-ray emission with the known location of these galaxies.  However the
signal to noise ratio of our data is insufficient to test this hypothesis
quantitatively.

\section{Summary and conclusions}

We have studied here a sample of 14 clusters of galaxies in a search for excess
soft X-ray emission.  In five clusters (Coma, A~1795, S\'ersic~159--03, MKW~3s
and A~2052) we detect excess soft X-ray emission, in four of these clusters
(NGC~533, A~496, A~262 and 2A~0335+096) excess absorption is found and in the
remaining five clusters no significant soft excess could be detected.

For three of the clusters with excess absorption this excess absorption can be
explained by enhanced foreground absorption related to the infrared cirrus.  For
the remaining cluster the compactness of its core in combination with steep
spectral gradients is the likely cause of less reliable fits.

In the five clusters with a detected soft X-ray excess, the strength of the
measured excess is well above the remaining calibration uncertainties of
XMM-Newton.

In at least three of these clusters (MKW~3s, A~2052 and S\'ersic~159--03) we
have firm evidence for the presence of extended emission (on an angular scale of
several degrees) due to warm gas of temperature $\sim$0.2~keV.  This extended
emission is also seen in maps of the ROSAT PSPC 1/4~keV band.  The surface
brightness of this spectral component on a cluster scale (within 12\arcmin\
radius) is more or less constant.  The thermal nature of this component is
supported by the detection of the unresolved \ion{O}{vii} triplet at the
redshift of the cluster.  We attribute this component to emission from the warm
intercluster medium.  Interestingly, in all these three clusters the ambient hot
cluster temperature drops rapidly in the region where the soft excess becomes
dominant.  Apparently we see here a smooth(ed) transition between the warm
intercluster medium and the hot intracluster medium.

In the other two clusters the same process can explain the soft excess emission,
although in the Coma cluster a power law model also gives a satisfactory fit,
and is consistent with the measured hard X-ray spectrum of this cluster.

Finally, in S\'ersic~159--03 there is an additional soft component that peaks
near the center of the cluster.  From the X-ray spectrum it is not fully
possible to distinguish different scenario's for this central component.
Possible scenario's include non-thermal X-ray radiation related to magnetic
reconnection, and thermal X-ray emission due to filaments filled with warm gas
and oriented along the line of sight towards the cluster.

\begin{acknowledgements}
We thank J.A.M.  Bleeker for allowing to use his Guaranteed Time data for this
analysis. We thank H. B\"ohringer for allowing to use the NGC~533 data for this
study. We thank an anonymous referee for helpfull remarks on the manuscript.
This work is based on observations obtained with XMM-Newton, an ESA science
mission with instruments and contributions directly funded by ESA Member States
and the USA (NASA).  SRON is supported financially by NWO, the Netherlands
Organization for Scientific Research.
\end{acknowledgements}

\end{document}